\documentclass[11pt,a4paper]{article}
\usepackage{jheppub,amsmath,amssymb,slashed,url,bm}
\usepackage{graphicx}
\usepackage{epstopdf}

\newcommand{\e}{\begin{eqnarray}}
\newcommand{\ee}{\end{eqnarray}}

\newcommand{\CN}{{\cal N}}
\newcommand{\CL}{{\cal L}}

\newcommand{\pa}{{a^{\prime}}}
\newcommand{\pb}{{b^{\prime}}}
\newcommand{\pc}{{c^{\prime}}}
\newcommand{\pd}{{d^{\prime}}}

\newcommand{\pu}{{u^{\prime}}}
\newcommand{\pv}{{v^{\prime}}}
\newcommand{\pw}{{w^{\prime}}}

\newcommand{\DA}{{\dot A}}
\newcommand{\DB}{{\dot B}}
\newcommand{\DC}{{\dot C}}
\newcommand{\DD}{{\dot D}}

\def\a{\alpha}
\def\b{\beta}
\def\d{\delta}
\newcommand{\ep}{\epsilon}
\newcommand{\g}{\gamma}
\newcommand{\p}{\psi}
\newcommand{\s}{\sigma}
\def\t{\tau}

\newcommand{\bp}{\bar{\psi}}
\newcommand{\BZ}{\bar{Z}}

\font\teneurm=eurm10 \font\seveneurm=eurm7  \font\fiveeurm=eurm5
\newfam\eurmfam
\textfont\eurmfam=\teneurm \scriptfont\eurmfam=\seveneurm
\scriptscriptfont\eurmfam=\fiveeurm

\font\teneusm=eusm10 \font\seveneusm=eusm7 \font\fiveeusm=eusm5
\newfam\eusmfam
\textfont\eusmfam=\teneusm \scriptfont\eusmfam=\seveneusm
\scriptscriptfont\eusmfam=\fiveeusm

\font\tencmmib=cmmib10 \skewchar\tencmmib='177
\font\sevencmmib=cmmib7 \skewchar\sevencmmib='177
\font\fivecmmib=cmmib5 \skewchar\fivecmmib='177
\newfam\cmmibfam
\textfont\cmmibfam=\tencmmib \scriptfont\cmmibfam=\sevencmmib
\scriptscriptfont\cmmibfam=\fivecmmib

\newcommand{\cj}{\mathcal{J}}

\title{$OSp(4|4)$ superconformal currents in
three-dimensional $\mathcal{N}=4$ Chern-Simons quiver gauge theories}

 \author{Fa-Min Chen }
\affiliation{Department of Physics, Beijing Jiaotong University, Beijing 100044, China}
\abstract{We prove explicitly that the general $D=3$, ${\cal N}=4$ Chern-Simons-matter (CSM) theory has a complete $OSp(4|4)$ superconformal symmetry, and construct the corresponding conserved currents.  We re-derive the $OSp(5|4)$ superconformal currents in the general $\CN=5$ theory as special cases of the $OSp(4|4)$ currents by enhancing the supersymmetry from $\CN=4$ to $\CN=5$. The closure of the full $OSp(4|4)$ superconformal algebra is verified explicitly.}
\begin{document} \maketitle

\section{Introduction and Summary} \label{Introduction}
The $D=3$, ${\cal N}=4$ Chern-Simons-matter (CSM)
theory was first constructed by Gaiotto and Witten (GW) \cite{GaWi}, by choosing the gauge groups carefully so that the ${\cal N}=1 $ supersymmetry can be promoted to ${\cal N}=4$. By adding twisted hyper-multiplets into the GW theory, the authors of Ref. \cite{HosomichiJD} have been able to construct an $\CN=4$ Chern-Simons quiver gauge theory.

The $\CN=4$ Chern-Simons quiver gauge theories are natural candidates of the dual gauge theories of multi M2-branes. For instance, the authors of \cite{HosomichiJD} constructed a class of $\CN=4$ theories with the closed loop quiver diagram of gauge groups (see also \cite{ChenWu6}):
\begin{equation}\label{linearN}
\cdots-U(N_{i-1})-U(N_i)-U(N_{i+1})-\cdots.
\end{equation}
(The above quiver diagram is only a part of the full diagram.) This special class of theories have been conjectured to be the dual gauge theories of multi
M2-branes in the orbifold $(\textbf{C}^2/\textbf{Z}_p\times \textbf{C}^2/\textbf{Z}_q)/\textbf{Z}_k$ \cite{Imamura}. Here $p$ and $q$ are the numbers of the un-twisted and twisted multiplets, respectively; $k$ is the Chern-Simons level. The corresponding gravity duals were studied in Ref. \cite{Imamura}.

By the gauge/gravity duality, the general $\CN=4$ theory is expected to have a full $OSp(4|4)$ superconformal symmetry. That is, the theory possesses an $\CN=4$ super Poincare symmetry as well as an $\CN=4$ superconformal symmetry\footnote{In this paper, the super Poincare transformations will be denoted as $\d_\ep$, satisfying $[\d_{\ep_1},\d_{\ep_2}]\sim P_\mu$, with $P_\mu$ the translations. The superconformal transformations will be denoted as $\d_\eta$, satisfying $[\d_{\eta_1},\d_{\eta_2}]\sim K_\mu$, with $K_\mu$ the special conformal transformations. The full super transformations (containing both $\d_\ep$ \emph{and} $\d_\eta$) will be called the $OSp(4|4)$ superconformal transformations.}. However, to our knowledge, only the law of the $\CN=4$ super Poincare transformations has been derived in the literature \cite{HosomichiJD}. (For a 3-algebra approach, see \cite{ChenWu3}.) To fill this gap, in this paper we derive the law of the $\CN=4$ superconformal transformations and verify the action is invariant under these transformations. We also derive the conserved supercurrents associated with the $\CN=4$ super Poincare transformations and the $\CN=4$ superconformal transformations. In other words, we prove that the $\CN=4$ theory possesses a complete $OSp(4|4)$ superconformal symmetry, and derive the full $OSp(4|4)$ superconformal currents.

We also demonstrate that the law of $OSp(5|4)$ superconformal transformations and the $OSp(5|4)$ superconformal currents \cite{Chen8} in the $\CN=5$ CSM theory, can be obtained as special cases of the law of the $OSp(4|4)$ superconformal transformations and the $OSp(4|4)$ currents of the $\CN=4$ theory, by enhancing the $SU(2)\times SU(2)$ R-symmetry to $USp(4)$. In our previous work \cite{Chen8}, we have showed that the $OSp(6|4)$ and $OSp(8|4)$ superconformal transformations and currents \cite{Schwarz,Schwarz0}, associated with the $\CN=6,8$ theories, respectively, can be derived as the special cases of the $OSp(5|4)$ superconformal transformations and currents\footnote{For a 3-algebra unifying $\CN=5,6,8$ theories, see \cite{Chen2,Chen:pku1,Palmkvist1}.}. Hence our approach provides a unified framework for all $\CN\geq4$ CMS theories.

To our best knowledge, so far only the closure of the $\CN=4$ super Poincare algebra of the $\CN=4$ theory has been checked (in the framework of 3-algebra) in the literature  \cite{ChenWu3}.
Therefore, it is necessary to verify the closure of the full $OSp(4|4)$ superconformal superalgebra. This is completed in the framework of Lie 2-algebra in Section \ref{SecClosure}.

The paper is organized as follows. In section \ref{secN4}, we derive the law of $\CN=4$ superconformal transformations and the corresponding conserved currents. In
Section \ref{secN5} we show that the $OSp(5|4)$ superconformal currents can be obtained as special cases of the $OSp(4|4)$ currents.
In Section \ref{SecClosure}, we check the closure of the full $OSp(4|4)$ superalgebra.
Our conventions and useful identities are summarized in Appendix \ref{Identities}. In Appendix \ref{SecN4Action}, we review the general $\CN=4$ theory. In Appendix \ref{SecDN4C}, we present the details of the derivation of the $\CN=4$ super Poincare currents.

\section{$OSp(4|4)$ Superconformal Currents}\label{secN4}
In this section we will construct the $\CN=4$ superconformal currents, and show that the $\CN=4$ theory has a full $OSp(4|4)$ superconformal symmetry. (The $\CN=4$ theory is reviewed in Appendix \ref{SecN4Action}, and our conventions are summarized in Appendix \ref{Identities}.)

The $\CN=4$ super Poincare currents are derived in Appendix \ref{SecDN4C}. They are given by
\e\label{superpcur2}
j^{I}_\mu&=&-i\bar\psi^A_\pa\g_\mu(\d\psi)^{I\pa}_A
-i\bar\psi^\DA_a\g_\mu(\d\psi)^{Ia}_\DA,
\ee
where
\e
(\d\psi)^{I\pa}_A&=&-\g^\mu D_\mu
Z^\pa_\DB\s^I_A{}^\DB-\frac{1}{3}k_{mn}\t^{m\pa}{}_{\pb}Z^\pb_\DB\mu^{\prime
n\DB}{}_\DC\s^I_A{}^\DC+k_{mn}\t^{m\pa}{}_\pb Z^\pb_\DA\mu^{nB}{}_A\s^I_B{}^\DA\\
(\d\psi)^{Ia}_\DA&=&-\g^\mu D_\mu
Z^a_B\s^{I\dag}_\DA{}^B-\frac{1}{3}k_{mn}\t^{ma}{}_{b}Z^b_B\mu^{
nB}{}_C\s^{I\dag}_\DA{}^C+k_{mn}\t^{ma}{}_b Z^b_A\mu^{\prime n\DB}{}_\DA\s^{I\dag}_\DB{}^A
\ee
are defined via the super Poincare transformations of the fermionic fields $\d_\ep\psi^\pa_A=(\d\psi)^{I\pa}_A\ep^I$ and $\d_\ep\psi^a_\DA=(\d\psi)^{Ia}_\DA\ep^I$ (see (\ref{2LN4})).
Here $A=1,2$, $\DA=1,2$, and $I=1,\ldots,4$ transform in the undotted, dotted, and vector representations of the $SU(2)\times SU(2)$ R-symmetry group, respectively; and $a=1,\ldots,2R$ and $\pa=1,\ldots, 2S$ transform in two different quaternionic representations of the gauge group. The scalar fields are denoted as $Z$. The sigma matrices $\s^I_A{}^\DB$ are defined in Appendix \ref{secSO4}.

Without changing the physical content, one can add a conserved total derivative term into the currents:
\e\label{newcurrent}
\tilde j^I_\mu=j^I_\mu+\partial^\nu A^I_{\mu\nu},
\ee
with $A^I_{\mu\nu}=-A^I_{\nu\mu}$, since the improved currents are still conserved, i.e. $\partial^\mu\tilde j^I_\mu=0$, and the set of super-charges  remain the same:
\e
Q^I=-\int d^2x\tilde j^I_0=-\int d^2x j^I_0.
\ee

The $\gamma$-trace $\g^\mu\tilde j^I_\mu$ measures the violation of scale invariance of the theory \cite{weinberg3}. Since the $\CN=4$ theory is invariant under scale transfromations, we expect that $\g^\mu\tilde j^I_\mu$ vanishes by choosing an appropriate $A^I_{\mu\nu}$. To see this, let us first calculate $\g^\mu j^I_\mu$:
\e
\g^\mu j^I_\mu=-i\partial^\nu[(Z^\pa_\DA\g_\nu\bar\psi^A_\pa+\bar Z^A_a\g_\nu\psi^a_\DA)\s^I_A{}^\DA].
\ee
This is equivalent to
\e\label{traceless}
\g^\mu [j^I_\mu+\frac{i}{4}[\g_\mu,\g_\nu]\partial^\nu(Z^\pa_\DA\bar\psi^A_\pa+\bar Z^A_a\psi^a_\DA)\s^I_A{}^\DA]=0.
\ee
This equation immediately suggests us to define
\e\label{superpcur3}
\tilde j^I_\mu=j^I_\mu+\frac{i}{4}[\g_\mu,\g_\nu]\partial
^\nu(Z^\pa_\DA\bar\psi^A_\pa+\bar Z^A_a\psi^a_\DA)\s^I_A{}^\DA.
\ee
As a result, the improved currents $\tilde j^I_\mu$ satisfy $\g^\mu\tilde j^I_\mu=0$.
In other words, Eq. (\ref{traceless}) suggests us to set $A^I_{\mu\nu}$ in (\ref{newcurrent}) as follows
\e
A^I_{\mu\nu}=\frac{i}{4}[\g_\mu,\g_\nu](Z^\pa_\DA\bar\psi^A_\pa+\bar Z^A_a\psi^a_\DA)\s^I_A{}^\DA.
\ee
Notice that $A^I_{\mu\nu}$ is indeed antisymmetric in $\mu$ and $\nu$.

Now it is possible to construct the new currents
\e
\tilde s^I_\mu=x\cdot\g\tilde j^I_\mu.
\ee
Using $\g^\mu\tilde j^I_\mu=0$ and $\partial^\mu \tilde j^I_\mu=0$, it is easy to prove that the new currents are also conserved: $\partial^\mu\tilde s^I_\mu=0$. The corresponding conserved supercharges are defined as follows:
\e
S^I=-\int d^2x\tilde s^I_0.
\ee
If we impose the equal-time commutators
\e\label{fmcommu}
&&\{\bar\psi^\DA_a(t,\vec{x}^\prime),\psi^b_\DB(t,\vec{x})\}
=-\d^\DA_\DB\d^b_a\gamma^0\d^2(x-x^\prime),\nonumber\\
&&\{\bar\psi^A_\pa(t,\vec{x}^\prime),\psi^\pb_B(t,\vec{x})\}
=-\d^A_B\d^\pb_\pa\gamma^0\d^2(x-x^\prime),\nonumber\\
&&[\Pi^A_a(t,\vec{x}^\prime),Z^b_B(t,\vec{x})]=-i\d^A_B\d^b_a\d^2(x-x^\prime)
,\nonumber\\
&&[\Pi^\DA_\pa(t,\vec{x}^\prime),Z^\pb_\DB(t,\vec{x})]=-i\d^\DA_\DB\d^\pb_\pa\d^2(x-x^\prime)
,\nonumber\\
&&[\Pi^\mu_m(t,\vec{x}^\prime),A^n_\nu(t,\vec{x})]=-i\d^n_m\d^\mu_\nu\d^2(x-x^\prime),
\ee
where $\Pi^A_a(t,\vec{x}^\prime)=D_0\bar Z^A_a(t,\vec{x}^\prime)$, $\Pi^\DA_\pa(t,\vec{x}^\prime)=D_0\bar Z^\DA_\pa(t,\vec{x}^\prime)$, and $\Pi^\mu_m(t,\vec{x}^\prime)=\ep^{\lambda 0\mu}k_{mp}A^p_\lambda(t,\vec{x}^\prime)$,
then the superconformal variantion of an arbitrary field $\Phi$ can be defined as
\e\label{sptrans}
\d_\eta\Phi=[-i\eta^IS^I,\Phi],
\ee
Using the above equation and the commutation relations (\ref{fmcommu}), one can readily derive the law of $\CN=4$ superconformal transformations (we will prove that $[\d_{\eta_1},\d_{\eta_2}]\sim K_\mu$ in Section \ref{SecClosure}):
\e \label{consusy}&&\delta_\eta Z^a_A=i(x\cdot\g\eta_A{}^\DA)\p^a_\DA,\nonumber\\
&&\d_\eta Z^\pa_\DA=i(x\cdot\g\eta^\dag_\DA{}^A)\p^\pa_A,\nonumber\\
&&\d_\eta\p^\pa_A=-\g^\mu D_\mu
Z^\pa_\DB(x\cdot\g\eta_A{}^\DB)-\frac{1}{3}k_{mn}\t^{m\pa}{}_{\pb}Z^\pb_\DB\mu^{\prime
n\DB}{}_\DC(x\cdot\g\eta_A{}^\DC)\nonumber\\&&\quad\quad\quad\quad+k_{mn}\t^{m\pa}{}_\pb Z^\pb_\DA\mu^{nB}{}_A(x\cdot\g\eta_B{}^\DA)-\eta_A{}^\DA Z^\pa_\DA, \nonumber\\
&&\d_\eta\p^a_\DA=-\g^\mu D_\mu
Z^a_B(x\cdot\g\eta^\dag_\DA{}^B)-\frac{1}{3}k_{mn}\t^{ma}{}_{b}Z^b_B\mu^{
nB}{}_C(x\cdot\g\eta^\dag_\DA{}^C)\nonumber\\&&\quad\quad\quad\quad+k_{mn}\t^{ma}{}_b Z^b_A\mu^{\prime n\DB}{}_\DA(x\cdot\g\eta^\dag_\DB{}^A)-\eta^\dag_\DA{}^AZ^a_A,\nonumber\\
&&\d_\eta A_\mu^m=i(x\cdot\g\eta^{A\DB})\g_\mu
j^m_{A\DB}+i(x\cdot\g\eta^{\dag\DA B})\g_\mu j^{\prime m}_{\DA B}.
\ee
The set of parameters $\eta_A{}^\DB=\eta^I\s^I{}_A{}^\DB$ $(I=1,\ldots,4)$ obey the reality conditions
\begin{equation}\label{n4para2}
\eta^{\dag}{}_{\dot{A}}{}^{B}= -\epsilon^{BC}\epsilon_{\dot{A}\dot{B}}\eta_{C}{}^{\dot{B}}.
\end{equation}

We now must prove that $\CN=4$ action (\ref{2LN4}) is invariant under the transformations (\ref{consusy}). Notice that the $\CN=4$ superconformal transformations (\ref{consusy}) can be obtained
by replacing $\ep^I$ by $x\cdot\g\eta^I$ in the $\CN=4$ super Poincare transformations (\ref{2SUSY4}) and adding the two additional terms \e\label{atranpsi}
\d^\prime_\eta\p^\pa_A=-\eta_A{}^\DA Z^\pa_\DA\;\;\;{\rm and}\;\;\;\d^\prime_{\eta}\psi^a_\DA=-\eta^\dag_\DA{}^AZ^a_A
\ee
into the transformations of the fermion fields $\p^\pa_A$ and $\psi^a_\DA$, respectively. (In a similar way, the $\CN=6,8$ superconformal transformations can be obtained from the $\CN=6,8$ super Poincare transformations \cite{Schwarz,Schwarz0}.) So the superconformal variation of the action can be calculated as follows:
\e\label{EdS5}
\d_\eta S&=&(\d S)_{\ep\rightarrow x\cdot\g\eta}+\d^\prime_\eta S\nonumber\\
&=&\int d^3x(-j^I_\mu)\partial^\mu(x\cdot\g\eta^I)+\d^\prime_\eta S,
\ee
where $(\d S)_{\ep\rightarrow x\cdot\g\eta}$ is the quantity obtained by replacing  $\ep^I$ in $\d_\ep S$ (see (\ref{EdS2})) by $x\cdot\g\eta^I$, and $\d^\prime_\eta S$ is the super variation of the action under the transformations (\ref{atranpsi}). A direct calculation gives
\e
\d^\prime_\eta S=-\int d^3x [(\g^\mu j^I_\mu)\eta^I].
\ee
Substituting the above equation into (\ref{EdS5}), we find that the action is indeed invariant under the transformations (\ref{consusy}), i.e.
\e
\d_\eta S=0.
\ee

\section{Unifying $OSp(4|4)$ and $OSp(5|4)$ Superconformal Currents}\label{secN5}

In this section we will demonstrate that the $OSp(5|4)$ superconformal currents, associated with the $\CN=5$ theory, can be derived as special cases of the $OSp(4|4)$ ones by enhancing the $SU(2)\times SU(2)$ R-symmetry to $USp(4)$.

In Ref. \cite{Hosomichi:2008jb}, it was showed that if the twisted and untwisted multiplets furnish the \emph{same} representations of the gauge group, i.e. if $\t^m_{ab}=\t^m_{\pa\pb}$, the $\CN=4$ supersymmetry will be enhanced to $\CN=5$ automatically. Specifically, if
\e\label{eqrep}
\t^m_{ab}=\t^m_{\pa\pb},
\ee
it is possible to embed the $SU(2)\times SU(2)$ R-symmetry group into $USp(4)$ by combining the $\CN=4$ twisted and untwisted multiplets to form the $\CN=5$ multiplets:
\begin{equation}\label{n5fields}
Z^a_A=
\begin{pmatrix}
 Z^a_A  \\
Z^a_\DA
\end{pmatrix},\quad
\psi^a_A=
\begin{pmatrix}
\psi^{a}_A  \\
\psi^a_{\DA}
\end{pmatrix}.
\end{equation}
In the left hand sides of the above equations, $A=1,\ldots,4$ transforms in the fundamental representation of $USp(4)$, while in the right hand sides, $A=1,2$ and $\DA=1,2$ transform in the undotted and dotted representations of $SU(2)\times SU(2)$. We hope this will not cause any confusion. In terms of these $\CN=5$ fields, the reality conditions (\ref{real}) become
\begin{eqnarray}\label{RealCondi}
\bar Z^A_a=\omega^{AB}\omega_{ab}Z^b_B ,\quad
\bar\psi^A_a=\omega^{AB}\omega_{ab}\psi^b_B,
\end{eqnarray}
where the antisymmetric tensor $\omega_{AB}$ of $USp(4)$ is defined as the charge conjugate matrix (see Appendix \ref{secSO5}):
\begin{equation}\label{antiform2}
\omega^{AB}=\begin{pmatrix} \epsilon^{AB} & 0 \\
0 & \epsilon^{\dot{A}\dot{B}}
\end{pmatrix}.
\end{equation}
In the RHS, $A$, $B$, $\dot A$, and $\dot B$ run from 1 to 2. Using (\ref{eqrep})$-$(\ref{antiform2}), the authors of Ref. \cite{Hosomichi:2008jb} have been able to promote the $\CN=4$ Lagrangian (\ref{2LN4}) to the manifestly $USp(4)$-invariant $\CN=5$ Lagrangian:
\begin{eqnarray}\label{5lagran}\nonumber
{\cal L}&=&\frac{1}{2}(-D_\mu\bar Z^A_aD^\mu
Z^a_A+i\bar\psi^A_a\gamma_\mu
D^\mu\psi^a_A)-\frac{i}{2}\omega^{AB}\omega^{CD}k_{mn}(\cj^m_{AC}\cj^n_{BD}-
2\cj^m_{AC}\cj^n_{DB})\nonumber\\
&&+\frac{1}{2}\epsilon^{\mu\nu\lambda}(k_{mn}A_\mu^m\partial_\nu
A_\lambda^n+\frac{1}{3}C_{mnp}A_\mu^mA_\nu^nA_\lambda^p)\\
&&+\frac{1}{30}C_{mnp}\mu^{mA}{}_B\mu^{nB}{}_C\mu^{pC}{}_A
+\frac{3}{20}k_{mp}k_{ns}(\t^m\t^n)_{ab}Z^{Aa}Z^b_A\mu^{pB}{}_C\mu^{sC}{}_B.\nonumber
\end{eqnarray}
The ``momentum map" and ``current" operators are defined as
\e\label{n5maps}
\mu^m_{AB}\equiv \t^m_{ab}Z^a_AZ^b_B, \quad
\mathcal{J}^m_{AB}\equiv\t^m_{ab}Z^a_A\p^b_B.\ee
The $\CN=5$ action is invariant under the $USp(4)$ R-symmetry transformations
\e\label{usp4tran}
\d_R\psi_{A}^a=\frac{1}{2}\ep_{IJ}\Sigma^{IJB}_A\psi_{B}^a,\quad
\d_RZ^a_{A}=\frac{1}{2}\ep_{IJ}\Sigma^{IJB}_AZ^a_{B},
\ee
where $\ep_{IJ}=-\ep_{JI}$ are set of parameters ($I,J=1,\ldots, 5$), and the $USp(4)$ generators $\Sigma^{IJA}_B$ are defined by (\ref{usp4ge}).

Using (\ref{eqrep})$-$(\ref{antiform2}), one can rewrite the $\CN=4$ super Poincare currents (\ref{superpcur3}) as
\e\label{superpcur4}
\tilde j^I_\mu=-i\bar\psi^A_a\g_\mu(\d\psi)^{Ia}_A+\frac{i}{4}[\g_\mu, \g_\nu]\partial^\nu(Z^a_A\bar\psi^B_a\Gamma^I_B{}^A), \quad (I=1,\ldots,4.)
\ee
where
\begin{equation}\label{svofp}
(\d\psi)^{Ia}_A\equiv-\gamma^{\mu}D_\mu Z^a_B\Gamma^I_A{}^B
-\frac{1}{3}k_{mn}\t^{ma}{}_{b}\omega^{BC}Z^b_B\mu^n_{CD}\Gamma^I_A{}^D
+\frac{2}{3}k_{mn}\t^{ma}{}_{b}\omega^{BD}Z^b_C\mu^n_{DA}\Gamma^I_B{}^C,
\end{equation}
with
\begin{eqnarray}\label{5Gamma2}
\Gamma^I_A{}^B=\begin{pmatrix}0&\sigma^I\\\sigma^{I\dag}&0
\end{pmatrix}\quad (I=1,\ldots, 4)
\end{eqnarray}
(See Appendix \ref{secSO5}.)
However, due to the $USp(4)$ R-symmetry, there is a \emph{fifth} conserved supercurrent. To see this, let consider for example the fourth supercurrent $\tilde j^4_\mu$. Under the $USp(4)$ R-symmetry transformations (\ref{usp4tran}), it becomes
\e
\tilde j^{4}_\mu&\rightarrow&\tilde j^{4}_\mu+\d_{R}\tilde j^{4}_\mu,\\
\d_R\tilde j^{4}_\mu&=&-\frac{1}{2}\ep_{KL}(\t^{KL})^{4}{}_J\tilde j^J_\mu, \quad (J, K, L=1,\ldots,5.)\label{conserved}
\ee
where $(\t^{KL})^{4}{}_J=\d^{L4}\d^K_J-\d^{K4}\d^L_J$ are a set of  $SO(5)$ matrices.
Since the action (\ref{5lagran}) is manifestly $USp(4)$-invariant, the transformed forth supercurrent must be conserved as well:
\e
\partial^\mu(\tilde j^{4}_\mu+\d_{R}\tilde j^{4}_\mu)=0.
\ee
This implies that
\e\label{conserved2}
\partial^\mu(\d_{R}\tilde j^{4}_\mu)=0.
\ee
Combining (\ref{conserved2}) and (\ref{conserved}) gives
\e
\partial^\mu\tilde j^I_\mu=0, \quad (I=1,\ldots,5)
\ee
where
\e\label{superpcur5}
\tilde j^I_\mu=-i\bar\psi^A_a\g_\mu(\d\psi)^{Ia}_A+\frac{i}{4}[\g_\mu, \g_\nu]\partial^\nu(Z^a_A\bar\psi^B_a\Gamma^I_B{}^A), \quad (I=1,\ldots,5.)
\ee
where $\Gamma^5$ is defined by equation (\ref{gamma5}).
In this way, we have obtained the $\CN=5$ super Poincare currents from the $\CN=4$ supercurrents. Similarly, the $\CN=5$ superconformal currents
\e\label{scu}
\tilde{s}^I_\mu=\g\cdot x\tilde{j}^I_\mu. \quad (I=1,\ldots,5)
\ee
can be derived from the $\CN=4$ superconformal currents.

In summary, our approach provides a unified framework for constructing the $OSp(4|4)$ and $OSp(5|4)$ superconformal currents. Furthermore, since the $OSp(6|4)$ and $OSp(8|4)$ superconformal currents of the $\CN=6$ and $\CN=8$ can be derived as the special cases of the $OSp(5|4)$ currents of the $\CN=5$ theory \cite{Chen8}, our approach actually provied a unified framework for constructing all conserved superconformal currents of the $\CN\ge4$ theories.

\section{Closure of the $OSp(4|4)$ Superconformal algebra}\label{SecClosure}

In the literature, only the closure of the super Poincare algebra of the $\CN=4$
Chern-Simons quiver gauge theory has been explicitly  checked (in the framework of 3-algebra) \cite{ChenWu3}.
In this section, we will verify the closure of the full $OSp(4|4)$ superconformal
superalgebra in the framework of Lie 2-algebra.

We begin by considering the scalar fields of the untwisted multiplets.
The commutator of two super Poincare transformations acting on the scalar fields gives \cite{ChenWu3}
\begin{eqnarray}\label{commZ1}
[\delta_{\ep_1}, \delta_{\ep_2}]Z^{a}_{A}=v_1^{\mu}D_{\mu}Z^{a}_{A}+\tilde{\Lambda}_1^a{}_bZ^b_A,
\end{eqnarray}
where\footnote{In our previous work \cite{ChenWu3}, we have examined the closure of the $\CN=4$ super Poincare algebra in the framework of 3-algebra. Here we convert the framework of 3-algebra into the conventional Lie algebra framework, using the method introduced in \cite{ChenWu6}.}
\begin{eqnarray}
v_1^{\mu} &=&-2i\ep_2^I\g^\mu\ep_1^I,\\
\tilde{\Lambda}_1^a{}_b&=&\Lambda^{m}_1k_{mn}\t^{na}{}_b, \\
\Lambda^{m}_1&=& -2i(\ep^I_1\ep^J_2)(\mu^m_{AB}\s^{IJAB}+\mu^{\prime m}_{\DA\DB}\bar\s^{IJ\DA\DB}).
\end{eqnarray}
Here $I=1,\ldots,4$ transforms in the fundamental representation of $SO(4)$, and the two
$SU(2)\times SU(2)$ matrices are defined as follows
\e\label{su2su2}
&&\s^{IJAB}=\frac{1}{4}(\s^I\s^{J\dag}-\s^J\s^{I\dag})^{AB},\quad
\bar\s^{IJ\DA\DB}=\frac{1}{4}(\s^{I\dag}\s^{J}-\s^{J\dag}\s^{I})^{\DA\DB}. \ee
The set of $SO(4)$ matrices $\s^I$ are defined in Appendix \ref{secSO4}.

Replacing $\ep_1^I$ in (\ref{commZ1}) by $x\cdot\g\eta^I_1$ and adding
\e\label{atranpsi1}
\d^\prime_{\eta_1}\psi^a_\DB=-\eta^\dag_{1\DB}{}^C Z^a_C
\ee
into the variation of the fermionic fields\footnote{The relation between $\d_\eta$ and $\d_\ep$ has been discussed in Section \ref{secN4}.}, we obtain
\begin{eqnarray}\label{commZ2}
[\delta_{\eta_1}, \delta_{\ep_2}]Z^{a}_{A}=v_2^{\mu}D_{\mu}Z^{a}_{A}+\tilde{\Lambda}_2^a{}_bZ^b_A-i\ep_{2A}{}^\DB\eta^\dag
_{1\DB}{}^CZ^a_C,
\end{eqnarray}
where
\e
v^\mu_2&=&-2i\ep_2^I\g^\mu(\g\cdot x\eta_1^I)\nonumber\\&=&-2i(\ep_2^I\eta^I_1)x^\mu+4(\ep^I_2\g^{\mu\nu}\eta^I_1)x_\nu,\\
\tilde{\Lambda}_2^a{}_b&=&\Lambda^{m}_2k_{mn}\t^{na}{}_b, \\
\Lambda^m_2&=&2ix^\mu(\ep^I_2\g_\mu\eta^J_1)(\mu^m_{AB}\s^{IJAB}+\mu^{\prime m}_{\DA\DB}\bar\s^{IJ\DA\DB}).
\ee
Notice that (\ref{commZ2}) is a gauge covariant equation, which can be simplified to give
\e
[\delta_{\eta_1}, \delta_{\ep_2}]Z^{a}_{A}&=&-2i(\eta^I_1\ep^I_2)(x^\mu\partial_\mu+\frac{1}{2})Z^a_A\nonumber\\
&&+
2(\eta_1^I\g^{\mu\nu}\ep^I_2)(x_\mu\partial_\nu-x_\nu\partial_\mu)Z^a_A\nonumber\\
&&+2i(\eta^I_1\ep^J_2)\s^{IJ}{}_A{}^BZ^a_B\nonumber\\
&&+(\tilde\Lambda_2^a{}_b+v^\mu_2A^m_\mu\t_m{}^a{}_b)Z^b_A.
\label{commZ4}
\ee
The first three lines are the scale, Lorentz, and R-symmetry transformations, respectively:
these transformations suggest that $\d_\eta$ is the superconformal transformation.
The last line is a gauge transformation. The first line indicates that the dimension of the scalar field is $\frac{1}{2}$, as expected.

Similarly, replacing $\ep_2^I$ in (\ref{commZ2}) by $x\cdot\g\eta^I_2$ and adding
\e\label{atranpsi2}
\d^\prime_{\eta_2}\psi^a_\DB=-\eta^\dag_{2\DB}{}^C Z^a_C
\ee
into the variation of the fermionic fields, we obtain
\begin{equation}\label{commZ3}
[\delta_{\eta_1}, \delta_{\eta_2}]Z^{a}_{A}=v_3^{\mu}D_{\mu}Z^{a}_{A}+
\tilde{\Lambda}_3^a{}_bZ^b_A-i(x\cdot\g\eta_{2A}{}^\DB)\eta^\dag
_{1\DB}{}^CZ^a_C+i(x\cdot\g\eta_{1A}{}^\DB)\eta^\dag
_{2\DB}{}^CZ^a_C,
\end{equation}
where
\e
v_3^{\mu} &=& -2i[(\g\cdot x\eta_2^I)\g^\mu(\g\cdot x\eta_1^I)]\label{v3}\nonumber
\\&=&-4i(\eta^I_1\g^\nu\eta^I_2)x_\nu x^\mu+2i(\eta^I_1\g^\mu\eta^I_2)x^2,\\
\tilde{\Lambda}_3^a{}_b&=&\Lambda^m_3k_{mn}\t^{na}{}_b,\label{lambda3} \\
\Lambda^m_3&=&-2i[(\g\cdot x\eta_2^I)(\g\cdot x\eta_1^J)](\mu^m_{AB}\s^{IJAB}+\mu^{\prime m}_{\DA\DB}\bar\s^{IJ\DA\DB})\nonumber\\&=&2ix^2(\eta^I_1\eta^J_2)(\mu^m_{AB}\s^{IJAB}+\mu^{\prime m}_{\DA\DB}\bar\s^{IJ\DA\DB}).
\ee
We observe that (\ref{commZ3}) is a gauge covariant equation. One can convert (\ref{commZ3}) into the form
\e\label{commZ5}
[\delta_{\eta_1}, \delta_{\eta_2}]Z^{a}_{A}&=&2(\eta^I_1\g^\nu\eta^I_2)[(-2ix_\nu x^\mu\partial_\mu+ix^2\partial_\nu)Z^a_A-ix_\nu Z^a_A]\nonumber\\&&+(\tilde\Lambda_3^a{}_b+v^\mu_3\tilde A_\mu^a{}_b)Z^b_A.
\ee
The first line is the standard special conformal variation: this again shows that $\d_\eta$ is indeed the superconformal transformation, as also suggested by Eq. (\ref{commZ4}). (The second line of (\ref{commZ5}) is a gauge transformation.)

Let us now consider the gauge fields \cite{ChenWu3}:
\begin{eqnarray}\label{A1}
[\delta_{\ep_1}, \delta_{\ep_2}]A^m_\mu&=& v_1^\nu F^m_{\nu\mu}-D_\mu\Lambda^m_1 \nonumber\\
&&+v_1^\nu\{F^m_{\mu\nu}-\varepsilon_{\mu\nu\lambda}[(Z^a_AD^\lambda \bar Z^{Ab}-\frac{i}{2}\bar{\psi}^{\DB a}\gamma^\lambda\psi^b_\DB)\t^m_{ab}\nonumber\\
&&\quad\quad\quad\quad+(Z^\pa_\DA D^\lambda \bar Z^{\DA\pb}-\frac{i}{2}\bar{\psi}^{B \pa}\gamma^\lambda\psi^\pb_B)\t^m_{\pa\pb}]\},
\end{eqnarray}
where the second and third lines are the equations of motion (EOM) for the gauge fields. Eq. (\ref{A1}) can be written as
\begin{eqnarray}\label{A2}
[\delta_{\ep_1}, \delta_{\ep_2}]A^m_\mu&=& v_1^\nu F^m_{\nu\mu}-(\ep_2^I\ep_1^J)D_\mu(\mu^m_{AB}\s^{IJAB}+\mu^{\prime m}_{\DA\DB}\bar\s^{IJ\DA\DB})+{\rm EOM}.
\end{eqnarray}
Applying the replacement $\ep_1\rightarrow x\cdot\g\eta_1$ to (\ref{A2}), and taking account of Eq. (\ref{atranpsi1}), we obtain the equation
\begin{eqnarray}\label{A3}
[\delta_{\eta_1}, \delta_{\ep_2}]A^m_\mu&=& v_2^\nu F^m_{\nu\mu}-D_\mu\Lambda^m_2 +{\rm EOM},
\end{eqnarray}
which can be converted into
\begin{eqnarray}\label{A4}
[\delta_{\eta_1}, \delta_{\ep_2}]A^m_\mu
&=&-2i(\eta^I_1\ep^I_2)(x^\mu\partial_\mu+1)A^m_\mu\nonumber\\&&+
2i(\eta_1^I\g^{\rho\sigma}\ep^I_2)[-i(x_\rho\partial_\sigma-x_\sigma\partial_\rho)A^m_\mu
+(S_{\rho\sigma})_\mu{}^\nu A^m_\nu]
\nonumber\\&&-D_\mu(\Lambda^m_2+v^\nu_2 A^m_\nu)\nonumber\\&& +{\rm EOM},
\end{eqnarray}
with $S_{\rho\sigma}$ the $SO(1,2)$ matrices:
\e\label{lorentz}
(S_{\rho\sigma})_\mu{}^\nu=i(\eta_{\sigma\mu}\d^\nu_\rho-
\eta_{\rho\mu}\d^\nu_\sigma).
\ee
It can be seen that the first three lines of (\ref{A4}) are the scale, Lorentz, and gauge transformations, respectively. From the first line we read off that the dimension of $A^m_\mu$ is $1$.

We now would like to evaluate $[\delta_{\eta_1}, \delta_{\eta_2}]A^m_\mu$. To do so, we first rewrite Eq. (\ref{A3}) as
\begin{equation}\label{A5}
[\delta_{\eta_1}, \delta_{\ep_2}]A^m_\mu=v_2^\nu F^m_{\nu\mu}-(\ep_2^I\g_\nu\eta_1^J)D_\mu[-2ix^\nu(\mu^m_{AB}\s^{IJAB}+\mu^{\prime m}_{\DA\DB}\bar\s^{IJ\DA\DB})] +{\rm EOM}
\end{equation}
Applying the replacement $\ep_2\rightarrow x\cdot\g\eta_2$ to (\ref{A5}), and using Eq. (\ref{atranpsi2}), we obtain the equation
\begin{eqnarray}\label{A6}
&&[\delta_{\eta_1}, \delta_{\eta_2}]A^m_\mu\nonumber\\
&=& v_3^\nu F^m_{\nu\mu}-D_\mu\Lambda^m_3\\
&=&2(\eta^I_1\g^\nu\eta^I_2)[(-2ix_\nu x^\rho\partial_\rho+ix^2\partial_\nu)A^m_\mu-2ix_\nu A^m_\mu-2x^\rho(S_{\rho\nu})_\mu{}^\sigma A^m_\sigma)]\label{A6}\nonumber\\&&
-D_\mu(\Lambda^m_3+v^\nu_3A^m_\mu)\label{A7}
\nonumber\\&& +{\rm EOM}.
\end{eqnarray}
The first line of (\ref{A7}) is indeed the special conformal transformation, while the second line is a gauge transformation.

Let us now recall the fermion supersymmetry transformation in Ref. \cite{ChenWu3}:
\begin{eqnarray}\label{4SusyOnPsi}
\nonumber [\delta_{\ep_1},\delta_{\ep_2}]\psi^a_{\DA} &=& v_1^\mu D_\mu
\psi^a_{\DA} + \tilde{\Lambda}_1^a{}_{b}
\psi^b_{\DA}\\
\nonumber &&-\frac{i}{2}(\epsilon_1^{\dag \DC B}\epsilon_{2B\DA}
-\epsilon_2^{\dag \DC B}\epsilon_{1B\DA})E^a_{\DC}\\
 &&
-\frac{1}{2}v_\nu\gamma^\nu E^a_{\DA},
\end{eqnarray}
where
\begin{equation}
0=E^a_{\DA} = \gamma^\mu D_\mu\psi^a_{\DA}
+\t_{m}{}^a{}_b(Z^b_Bj^{mB}{}_\DA-\mu^{\prime m}_{\DA\DC}\psi^{\DC b}+2Z^b_Bj^{\prime m}{}_\DA{}^B)
\end{equation}
are equations of motion for the fermionic fields.

Using the same trick for evaluating the scalar and gauge fields, we obtain
\begin{eqnarray}\label{SusyOnPsi2}
\nonumber [\delta_{\eta_1},\delta_{\ep_2}]\psi^a_{\DA} &=& [v_2^\mu D_\mu
\psi^a_{\DA} -i(\psi^a_\DC\g^\mu\eta_{1B}{}^\DC)(\g_\mu\ep^\dag_{2\DA}{}^B)]
+i\eta^\dag_{1\DA}{}^B(\ep_{2B}{}^\DC\psi^a_\DC)
\nonumber\\&&+ \tilde{\Lambda}_2^a{}_{b}
\psi^b_{\DA}
\\&&+ {\rm EOM}\nonumber
\end{eqnarray}
Using the Fierz transformations (\ref{Fierz}), one can recast the above equation as
\begin{eqnarray}
[\delta_{\eta_1},\delta_{\ep_2}]\psi^a_{\DA}&=&-2i(\eta^I_1\ep^I_2)(x^\mu\partial_\mu+1)\psi^a_\DA
\nonumber\\&&+
2(\eta_1^I\g^{\mu\nu}\ep^I_2)(x_\mu\partial_\nu-x_\nu\partial_\mu
+i\g_{\mu\nu})\psi^a_\DA\nonumber\\
&&+2i(\eta^I_1\ep^J_2)\bar\s^{IJ}{}_\DA{}^\DB\psi^a_\DB\nonumber\\&&
+(\tilde\Lambda_2^a{}_b+v^\mu_2A^m_\mu\t_m{}^a{}_b)\psi^b_\DA
\nonumber\\&&+{\rm EOM}.
\end{eqnarray}
The first three lines are the scale, Lorentz, and R-symmetry transformations, while the fourth line is a gauge transformation. The first line indicates that the dimension of the spinor field is $1$.

The commutator of two superconformal transformations acting on the fermionic fields
gives
\e\label{psi4}
&&[\delta_{\eta_1}, \delta_{\eta_2}]\psi^{a}_{\DA}\nonumber\\
&=&2(\eta^I_1\g^\nu\eta^I_2)[(-2ix_\nu x^\mu\partial_\mu+ix^2\partial_\nu)Z^a_A+(-2ix_\nu -2x^\mu\g_{\mu\nu})\psi^a_\DA]\nonumber\\&&+(\tilde\Lambda_3^a{}_b+v^\mu_3\tilde A_\mu^a{}_b)\psi^b_\DA\nonumber\\&&+{\rm EOM}.
\ee
The first line is nothing but the special conformal transformation of the fermionic fields, and the second line is a gauge transformation.
Also, Eq. (\ref{psi4}) can recast as the following gauge invariant form:
\e\label{psi5}
&&[\delta_{\eta_1}, \delta_{\eta_2}]\psi^{a}_{\DA}\nonumber\\
&=&2(\eta^I_1\g^\nu\eta^I_2)[(-2ix_\nu x^\mu D_\mu+ix^2D_\nu)Z^a_A+(-2ix_\nu -2x^\mu\g_{\mu\nu})\psi^a_\DA]
\nonumber\\&&+\tilde\Lambda_3^a{}_b\psi^b_\DA+{\rm EOM}.
\ee

The transformations of the scalar and spinor fields of the twisted multiplets have similar expressions of that of the untwisted multiplets. First of all, the $\CN=4$ super Poincare algebra is closed (on-shell) on the twisted multiplets \cite{ChenWu3}:
\e\label{pz1}
&&[\delta_{\ep_1}, \delta_{\ep_2}]Z_\DA^\pa=v^\mu_1 D_\mu
Z^\pa_\DA+\tilde\Lambda_1^\pa{}_\pb Z^\pb_\DA,\\
&&[\delta_{\ep_1},\delta_{\ep_2}]\psi^\pa_{A} = v_1^\mu D_\mu
\psi^\pa_{A} + \tilde{\Lambda}_1^\pa{}_{\pb} \psi^\pb_{A}+{\rm EOM},\label{ppsi1}
\ee
where $\Lambda_1^\pa{}_\pb=k_{mn}\Lambda^m_1\t^{n\pa}{}_\pb$, and the equations of motion for the spinor fields are given by
\begin{equation}
0=E^\pa_{A} = \gamma^\mu D_\mu\psi^\pa_{A}
+\t_{m}{}^\pa{}_\pb(Z^\pb_\DB j^{\prime m\DB}{}_A-\mu^{m}_{AC}\psi^{C \pb}+2Z^\pb_\DB j^{m}{}_A{}^\DB)
\end{equation}

Secondly, taking advantage of the strategy for calculating the transformations of the fields of the untwisted multiplets, we find that the rest of the $OSp(4|4)$ superalgebra is also closed (on-shell) on the scalar and fermionic fields of the twisted multiplets:
\e
[\delta_{\eta_1}, \delta_{\ep_2}]Z^{\pa}_{\DA}&=&-2i(\eta^I_1\ep^I_2)
(x^\mu\partial_\mu+\frac{1}{2})Z^\pa_\DA\nonumber\\
&&+
2(\eta_1^I\g^{\mu\nu}\ep^I_2)(x_\mu\partial_\nu-x_\nu\partial_\mu)Z^\pa_\DA\nonumber\\
&&+2i(\eta^I_1\ep^J_2)\bar\s^{IJ}{}_\DA{}^\DB Z^\pa_\DB\nonumber\\
&&+(\tilde\Lambda_2^\pa{}_\pb+v^\mu_2A^m_\mu\t_m{}^\pa{}_\pb)Z^\pb_\DA,
\label{commpZ4}
\ee
\begin{eqnarray}
[\delta_{\eta_1},\delta_{\ep_2}]\psi^\pa_{A}&=&-2i(\eta^I_1\ep^I_2)(x^\mu\partial_\mu+1)\psi^\pa_A
\nonumber\\&&+
2(\eta_1^I\g^{\mu\nu}\ep^I_2)(x_\mu\partial_\nu-x_\nu\partial_\mu
+i\g_{\mu\nu})\psi^\pa_A\nonumber\\
&&+2i(\eta^I_1\ep^J_2)\s^{IJ}{}_A{}^B\psi^\pa_B\nonumber\\&&
+(\tilde\Lambda_2^\pa{}_\pb+v^\mu_2A^m_\mu\t_m{}^\pa{}_\pb)\psi^\pb_A
\nonumber\\&&+{\rm EOM},
\end{eqnarray}
\e\label{commpZ5}
[\delta_{\eta_1}, \delta_{\eta_2}]Z^{\pa}_{\DA}&=&2(\eta^I_1\g^\nu\eta^I_2)[(-2ix_\nu x^\mu\partial_\mu+ix^2\partial_\nu)Z^\pa_\DA-ix_\nu Z^\pa_\DA]\nonumber\\&&+(\tilde\Lambda_3^\pa{}_\pb+v^\mu_3\tilde A_\mu^\pa{}_\pb)Z^\pb_\DA,
\ee
and
\e\label{ppsi5}
&&[\delta_{\eta_1}, \delta_{\eta_2}]\psi^{\pa}_{A}\nonumber\\
&=&2(\eta^I_1\g^\nu\eta^I_2)[(-2ix_\nu x^\mu D_\mu+ix^2D_\nu)Z^\pa_\DA+(-2ix_\nu -2x^\mu\g_{\mu\nu})\psi^\pa_A]
\nonumber\\&&+\tilde\Lambda_3^\pa{}_\pb\psi^\pb_A+{\rm EOM}.
\ee


\section{Acknowledgement}
We thank the referee for useful comments. This work is supported by the China Postdoctoral Science Foundation through Grant No. 2012M510244.

\appendix

\section{Conventions and Useful Identities}\label{Identities}

The conventions and identities of this appendix are  adopted from Ref. \cite{Chen8}.
\subsection{Spinor Algebra}
In $1+2$ dimensions, the gamma matrices are defined as
\begin{equation}
(\gamma_{\mu})_{\alpha}{}^\gamma(\gamma_{\nu})_{\gamma}{}^\beta+
(\gamma_{\nu})_{\alpha}{}^\gamma(\gamma_{\mu})_{\gamma}{}^\beta=
2\eta_{\mu\nu}\delta_{\alpha}{}^\beta.
\end{equation} For the metric we
use the $(-,+,+)$ convention. The gamma matrices in the Majorana
representation can be defined in terms of Pauli matrices:
$(\gamma_{\mu})_{\alpha}{}^\beta=(i\sigma_2, \sigma_1, \sigma_3)$,
satisfying the important identity
\begin{equation}
(\gamma_{\mu})_{\alpha}{}^\gamma(\gamma_{\nu})_{\gamma}{}^\beta
=\eta_{\mu\nu}\delta_{\alpha}{}^\beta+\varepsilon_{\mu\nu\lambda}(\gamma^{\lambda})_{\alpha}{}^\beta.
\end{equation}
We also define
$\varepsilon^{\mu\nu\lambda}=-\varepsilon_{\mu\nu\lambda}$. So
$\varepsilon_{\mu\nu\lambda}\varepsilon^{\rho\nu\lambda} =
-2\delta_\mu{}^\rho$. We raise and lower spinor indices with an
antisymmetric matrix
$\epsilon_{\alpha\beta}=-\epsilon^{\alpha\beta}$, with
$\epsilon_{12}=-1$. For example,
$\psi^\alpha=\epsilon^{\alpha\beta}\psi_\beta$ and
$\gamma^\mu_{\alpha\beta}=\epsilon_{\beta\gamma}(\gamma^\mu)_\alpha{}^\gamma
$, where $\psi_\beta$ is a Majorana spinor. Notice that
$\gamma^\mu_{\alpha\beta}=(\mathbb{I}, -\sigma^3, \sigma^1)$ are
symmetric in $\alpha$ and $\beta$. A vector can be represented by a
symmetric bispinor and vice versa:
\begin{equation}
A_{\alpha\beta}=A_\mu\gamma^\mu_{\alpha\beta},\quad\quad A_\mu=-\frac{1}{2}\gamma^{\alpha\beta}_\mu A_{\alpha\beta}.
\end{equation}
We use the following spinor summation convention:
\begin{equation}
\psi\chi=\psi^\alpha\chi_\alpha,\quad\quad
\psi\gamma_\mu\chi=\psi^\alpha(\gamma_{\mu})_{\alpha}{}^\beta\chi_\beta,
\end{equation}
where $\psi$ and $\chi$ are anti-commuting Majorana spinors. In
$1+2$ dimensions the Fierz transformations are
\begin{eqnarray}\label{Fierz}
(\lambda\chi)\psi &=& -\frac{1}{2}(\lambda\psi)\chi -\frac{1}{2}
(\lambda\gamma_\nu\psi)\gamma^\nu\chi.
\end{eqnarray}
There is another useful identity:
\e\label{totantiS2}
(\psi_1\psi_2)\psi_3+(\psi_2\psi_3)\psi_1+(\psi_3\psi_1)\psi_2=0,
\ee
where $\psi_1$, $\psi_2$, and $\psi_3$ are arbitrary spinors.
Finally, we define
\e\label{gammamunu1}
\g^{\mu\nu}=-\frac{i}{4}[\g^\mu, \g^\nu].
\ee

\subsection{$SO(4)$ and $SO(5)$ Gamma Matrices}\label{SecSOM}
\subsubsection{$SO(4)$ Gamma Matrices}\label{secSO4}
We define the 4 sigma matrices as
\begin{equation}\label{pulim}
\sigma^I{}_A{}^{\dot{B}}=(\sigma^1,\sigma^2,\sigma^3,i\mathbb{I}),
\end{equation}
by which one can establish a connection between the $SU(2)\times
SU(2)$ and $SO(4)$ group. These sigma matrices satisfy the following
Clifford algebra:
\begin{eqnarray}
\sigma^I{}_{A}{}^{\dot{C}}\sigma^{J\dag}{}_{\dot{C}}{}^B+
\sigma^J{}_{A}{}^{\dot{C}}\sigma^{I\dag}{}_{\dot{C}}{}^B=2\delta^{IJ}\delta_A{}^B,\\
\sigma^{I\dag}{}_{\dot{A}}{}^{C}\sigma^{J}{}_{C}{}^{\dot{B}}+
\sigma^{J\dag}{}_{\dot{A}}{}^{C}\sigma^{I}{}_{C}{}^{\dot{B}}=2\delta^{IJ}\delta_{\dot{A}}{}^{\dot{B}}.
\end{eqnarray}
We use antisymmetric matrices
\begin{eqnarray}
\epsilon_{AB}=-\epsilon^{AB}=\begin{pmatrix} 0&-1 \\ 1&0
\end{pmatrix}\;
\;\;{\rm and}\;\;\;
\epsilon_{\dot{A}\dot{B}}=-\epsilon^{\dot{A}\dot{B}}=\begin{pmatrix}
0&1 \\-1& 0
\end{pmatrix}
\end{eqnarray}
to raise or lower un-dotted and dotted indices, respectively. For
example,
$\sigma^{I\dag\dot{A}B}=\epsilon^{\dot{A}\dot{B}}\sigma^{I\dag}{}_{\dot{B}}{}^{B}$
and $\sigma^{IB\dot{A}}=\epsilon^{BC}\sigma^{I}{}_{C}{}^{\dot{A}}$.
The sigma matrices $\sigma^I$ satisfy the reality conditions
\begin{equation}\label{RC4}
\sigma^{I\dag}{}_{\dot{A}}{}^{B}=-\epsilon^{BC}\epsilon_{\dot{A}\dot{B}}\sigma^I{}_{C}{}^{\dot{B}},\quad
{\rm or} \quad\sigma^{I\dag\dot{A}B}=-\sigma^{IB\dot{A}}.
\end{equation}
The antisymmetric matrices $\epsilon_{AB}$ and $\epsilon_{\dot{A}\dot{B}}$ satisfy the
identities
\e
\epsilon_{AB}\epsilon^{CD}&=&-(\delta_A{}^C\delta_B{}^{D}-\delta_A{}^D\delta_B{}^{C}),\\
\epsilon_{\DA\DB}\epsilon^{\DC\DD}&=&-(\delta_\DA{}^\DC\delta_\DB{}^{\DD}-\delta_\DA{}^\DD\delta_\DB{}^{\DC}).
\ee

\subsubsection{$SO(5)$ Gamma Matrices}\label{secSO5}

Using (\ref{pulim}), we define the first four $SO(5)$ gamma matrices as \footnote{To avoid introducing too many indices, we still use the capital letters $A, B, \ldots$ to
label the $USp(4)$ indices, and use $I$ to label a fundamental index of $SO(4)$. We hope this will not cause any confusion.}
\begin{eqnarray}\label{5Gamma}
\Gamma^I_A{}^B=\begin{pmatrix}0&\sigma^I\\\sigma^{I\dag}&0
\end{pmatrix}\quad (I=1,\ldots, 4),
\end{eqnarray}
and define the fifth $SO(5)$ gamma matrix as
\e\label{gamma5}
\Gamma^5_A{}^B=(\Gamma^1\Gamma^2\Gamma^3\Gamma^4)_A{}^B.
\ee
Notice that
$\Gamma^I_A{}^B$ ($I=1,\ldots, 5$) are Hermitian, satisfying the
Clifford algebra
\begin{equation}\label{so5g}
\Gamma^{I}_A{}^C\Gamma^{J}_C{}^B+\Gamma^{J}_A{}^C\Gamma^{I}_C{}^B
=2\delta^{IJ}\delta_A{}^B.
\end{equation}
We use an antisymmetric matrix $\omega_{AB}=-\omega^{AB}$ to lower
and raise indices; for instance
\begin{equation}\label{5raise}
\Gamma^{IAB}=\omega^{AC}\Gamma^I_C{}^B.
\end{equation}
It can be chosen as the charge conjugate matrix:
\begin{equation}\label{antiform}
\omega^{AB}=\begin{pmatrix} \epsilon^{AB} & 0 \\
0 & \epsilon^{\dot{A}\dot{B}}
\end{pmatrix}.
\end{equation}
(Recall that $A$ and $\dot{B}$ of the RHS run from 1 to 2.)

By the definition (\ref{5Gamma}) and the convention (\ref{5raise}),
the gamma matrix $\Gamma^I$ is antisymmetric and traceless, and
satisfies a reality condition
\begin{eqnarray}
\Gamma^{IAB}=-\Gamma^{IBA} \quad,\quad \Gamma^{I}_A{}^A=0\quad {\rm
and}\quad
\Gamma^{I*}_{AB}=\Gamma^{IAB}=\omega^{AC}\omega^{BD}\Gamma^I_{CD}.
\end{eqnarray}
The $USp(4)$ generators are defined as
\begin{equation}\label{usp4ge}
\Sigma^{IJ}_A{}^B=\frac{1}{4}[\Gamma^I, \Gamma^J]_A{}^B.
\end{equation}


\section{A review of the $\CN=4$ theory }\label{SecN4Action}
In this appendix, we review the general $\CN=4$ theory constructed in \cite{HosomichiJD}. This theory was constructed by generalizing the $\CN=4$ GW theory \cite{GaWi} to include twisted multiplets. So the theory contains both twisted and un-twisted multiplets. Both the twisted and untwisted multiplets are required to furnish quaternoinic representations of the gauge group. Generally speaking, the quaternoinic representation furnished by the twisted multiplets is \emph{not} equivalent to the representation furnished by the untwisted multiplets.

We denote the untwisted and twisted multiplets by $(Z^a_A, \p^a_\DA)$ and $(Z^\pa_\DA, \p^\pa_A)$, respectively. Here $A,\DA=1,2$ transform in the two-dimensional representation of the $SU(2)\times SU(2)$ R-symmetry group; $a=1,\ldots,2R$ transforms in a quaternionic representation of the gauge group, and $\pa=1,\ldots, 2S$ transforms in \emph{another} quaternionic representation of the gauge group. The corresponding quaternionic forms are denoted as $\omega_{ab}$ and $\omega_{\pa\pb}$,  satisfying
$\omega_{ab}\omega^{bc}=\delta^c_a$ and $\omega_{\pa\pb}\omega^{\pb\pc}=\delta^\pc_\pa$. We use the antisymmetric tensors $\omega$ to lower or raise the indices; for instance, $\t^m_{ab}=\omega_{ac}\t^{mc}{}_b$, where $\t^{mc}{}_b$ are a set of representation matrices of the Lie algebra of gauge symmetry. The
un-twisted multiplets $(Z^a_A, \p^a_\DA)$ and twisted
multiplets $(Z^\pa_\DA, \p^\pa_A)$ satisfy the reality
conditions:
\e\label{real}
\bar Z^A_a&=&\omega_{ab}\ep^{AB}Z^b_B,\quad \bar
\p^\DA_a=\omega_{ab}\ep^{\DA\DB}Z^b_\DB,\nonumber\\
\bar Z^\DA_\pa&=&\omega_{\pa\pb}\ep^{\DA\DB}Z^\pb_\DB,\quad \bar
\p^A_\pa=\omega_{\pa\pb}\ep^{AB}Z^\pb_B,
\ee
where $\ep^{AB}$ and $\ep^{\DA\DB}$ are antisymmetric forms of the $SU(2)\times SU(2)$ R-symmetry group (see Appendix \ref{secSO4}).

To be compatible with the $\CN=4$ supersymmetry,  the representation matrices $\t^m_{ab}$ and $\t^m_{\pa\pb}$ of the gauge group are required to satisfy the fundamental identities
\e\label{FI}
k_{mn}\t^m_{(ab}\t^n_{c)d}=0,\quad k_{mn}\t^m_{(\pa\pb}\t^n_{\pc)\pd}=0.
\ee

Following Ref. \cite{GaWi}, we define the ``momentum maps" and
``currents" as
\e\label{maps} \mu^m_{AB}\equiv \t^m_{ab}Z^a_AZ^b_B, \quad
j^m_{A\DB}\equiv\t^m_{ab}Z^a_A\p^b_\DB,\quad \mu^{\prime
m}_{\DA\DB}\equiv \t^m_{\pa\pb}Z^\pa_\DA Z^\pb_\DB, \quad j^{\prime
m}_{\DA B}\equiv\t^m_{\pa\pb}Z^\pa_\DA\p^\pb_B.\ee
Denoting the gauge fields as $A^m_\mu$, the
Lagrangian of the $\CN=4$ CSM theory reads \cite{HosomichiJD,ChenWu6}
\e\label{2LN4}
\CL&=&\frac{1}{2}\epsilon^{\mu\nu\lambda}(k_{mn}A_\mu^m\partial_\nu
A_\lambda^n+\frac{1}{3}C_{mnp}A_\mu^mA_\nu^nA_\lambda^p)\nonumber\\
&&+\frac{1}{2}(-D_\mu\bar{Z}^A_aD^\mu Z^a_A-D_\mu\bar{Z}^\DA_\pa
D^\mu Z^\pa_\DA+i\bp^\DA_a\g^\mu D_\mu\p^a_\DA+i\bp^A_\pa\g^\mu
D_\mu\p^\pa_A)\nonumber\\&&-\frac{i}{2}k_{mn}(j^m_{A\DB}j^{nA\DB}
+j^{\prime m}_{\DA B}j^{\prime n\DA B}-4j^m_{A\DB}j^{\prime n\DB
A})\nonumber\\&&+\frac{i}{2}k_{mn}(\mu^m_{AB}\t^n_{\pa\pb}\p^{A\pa}\p^{B\pb}
+\mu^{\prime m}_{\DA\DB}\t^n_{ab}\p^{\DA a}\p^{\DB b})\\&&
-\frac{1}{24}C_{mnp}(\mu^{mA}{}_B\mu^{nB}{}_C\mu^{pC}{}_A
+\mu^{\prime m\DA}{}_\DB\mu^{\prime n\DB}{}_\DC\mu^{\prime
p\DC}{}_\DA)\nonumber\\&&+\frac{1}{4}k_{mp}k_{ns}((\t^m\t^n)_{ab}Z^{Aa}Z^b_A\mu^{\prime
p\DB}{}_\DC\mu^{\prime
s\DC}{}_\DB+(\t^m\t^n)_{\pa\pb}Z^{\DA\pa}Z^\pb_\DA\mu^{ pB}{}_C\mu^{
sC}{}_B)\nonumber. \ee
We have used the invariant form $k_{ms}$ on the Lie algebra of the gauge
group to lower the indices of the structure
constants: $C_{mnp}=k_{ms}k_{nq}C^{sq}{}_p$.
(We will also define $\t_{m}{}^a{}_b=k_{mn}\t^{na}{}_b$.)

The $\CN=4$ super Poincare
transformations are given by
\e \label{2SUSY4}&&\delta_\ep Z^a_A=i\ep_A{}^\DA\p^a_\DA,\nonumber\\
&&\d_\ep Z^\pa_\DA=i\ep^\dag_\DA{}^A\p^\pa_A,\nonumber\\
&&\d_\ep\p^\pa_A=-\g^\mu D_\mu
Z^\pa_\DB\ep_A{}^\DB-\frac{1}{3}k_{mn}\t^{m\pa}{}_{\pb}Z^\pb_\DB\mu^{\prime
n\DB}{}_\DC\ep_A{}^\DC+k_{mn}\t^{m\pa}{}_\pb Z^\pb_\DA\mu^{nB}{}_A\ep_B{}^\DA, \nonumber\\
&&\d_\ep\p^a_\DA=-\g^\mu D_\mu
Z^a_B\ep^\dag_\DA{}^B-\frac{1}{3}k_{mn}\t^{ma}{}_{b}Z^b_B\mu^{
nB}{}_C\ep^\dag_\DA{}^C+k_{mn}\t^{ma}{}_b Z^b_A\mu^{\prime n\DB}{}_\DA\ep^\dag_\DB{}^A,\nonumber\\
&&\d_\ep A_\mu^m=i\ep^{A\DB}\g_\mu
j^m_{A\DB}+i\ep^{\dag\DA B}\g_\mu j^{\prime m}_{\DA B}.
\ee
Here the parameters $\ep_A{}^\DB=\ep^I\s^I{}_A{}^\DB$ $(I=1,\ldots,4)$ obey the reality conditions
\begin{equation}\label{n4para}
\ep^{\dag}{}_{\dot{A}}{}^{B}= -\epsilon^{BC}\epsilon_{\dot{A}\dot{B}}\ep_{C}{}^{\dot{B}}.
\end{equation}

The fundamental identities (\ref{FI}) can be converted into certain Jacobi identities of two superalgebras $G$ and $G^\prime$ admitting quaternoinic structures \cite{GaWi}; and the Lie algebra of the gauge symmetry is the bosonic parts of $G$ \emph{and} $G^\prime$. More concretely ,
the Lie algebra of the gauge group is given by the bosonic subalgebra of the superalgebra $G$ \cite{GaWi}
\begin{equation}\label{slie2} [M^u, M^v]=f^{uv}{}_wM^w,\quad [M^u,
Q_a]=-\t^u_{ab}\omega^{bc}Q_c,\quad \{Q_a,Q_b\}=\t^u_{ab}k_{uv}M^v,
\end{equation}
\emph{and} the bosonic subalgebra of the superalgebra $G^\prime$
\begin{equation}\label{slie3} [M^\pu, M^\pv]=f^{\pu\pv}{}_\pw M^\pw,\quad [M^\pu,
Q_\pa]=-\t^\pu_{\pa\pb}\omega^{\pb\pc}Q_\pc,\quad
\{Q_\pa,Q_\pb\}=\t^\pu_{\pa\pb}k_{\pu\pv}M^\pv.
\end{equation}
In order that there are physical interactions between the twisted and untwisted multiplets, the bosonic parts of the superalgebras $G$ and $G^\prime$ are required to share at least one simple factor or $U(1)$ factor \cite{HosomichiJD,ChenWu6}. More precisely, decomposing $M^u$ and $M^\pu$ as
$M^u=(M^\a, M^g)$ and $M^u=(M^{\a^\prime}, M^g)$, with $M^g$ the common generators, then the Lie algebra of the gauge symmetry is spanned by the generators \cite{HosomichiJD,ChenWu6}
\e
M^m=(M^\a,M^g,M^{\a^\prime}).
\ee
In accordance with the decompositions of $M^u$ and $M^\pu$ , we may decompose the quadratic forms and structure constants as follows
\e
&&k_{uv}=(k_{\a\b}, k_{gh}),\quad k_{\pu\pv}=(k_{\a^\prime\b^\prime}, k_{gh}),\\
&&f^{uv}{}_w=(f^{\a\b}{}_\g,f^{fg}{}_h),\quad f^{\pu\pv}{}_\pw=(f^{\a^\prime\b^\prime}{}_{\g^\prime},f^{fg}{}_h).
\ee

Let us now put (\ref{slie2}) \emph{and}
(\ref{slie3}) together: \e\label{Fsdspalg2}
&&[M^m,M^n]=C^{mn}{}_p M^p,\quad [M^m,
Q_a]=-\t^m_{ab}\omega^{bc}Q_c,\quad [M^m,
Q_\pa]=-\t^m_{\pa\pb}\omega^{\pb\pc}Q_\pc,\nonumber\\
&& \{Q_{a},Q_{b}\}=\t^m_{ab}k_{mn}M^n, \quad \{Q_{\pa},Q_{\pb}\}=\t^m_{\pa\pb}k_{mn}M^n,
\ee
%
where
\begin{eqnarray}\label{drcsm1}
&&C^{mn}{}_p=(f^{\a\b}{}_\g,
f^{fg}{}_h,f^{\a^\prime\b^\prime}{}_{\g^\prime}),\\
&&k_{mn}=(k_{\a\b},k_{gh},k_{\a^\prime\b^\prime}).\label{drcsm3}
\end{eqnarray}
It can be seen that the fundamental identities (\ref{FI}) are equivalent to the $Q_aQ_bQ_c$ and $Q_\pa Q_\pb Q_\pc$ Jacobi identities of (\ref{drcsm1}) \cite{GaWi}.
The classification of the gauge groups can be found in \cite{HosomichiJD,ChenWu6}.

If the twisted and untwisted multiplets form the \emph{same} representation of gauge group\footnote{This can be achieved by letting both twisted and untwisted multiplets to take the representation of the bosonic subalgebra of $G$ (\ref{slie2}); this representation is furnished by the set of fermionic generators $Q_a$.}, the $\CN=4$ supersymmetry can be enhanced to $\CN=5$ \cite{Hosomichi:2008jb}.

\section{Derivation of the $\CN=4$ Super Poincare Currents}\label{SecDN4C}
In this Appendix, we will derive the $\CN=4$ super Poincare currents by using the standard Noether method.

The super Poincare variation of the action (\ref{2LN4}) must take the form
\e\label{EdS2}
\d_\ep S=\int d^3x(-j^I_\mu)\partial^\mu\ep^I,
\ee
if we allow the set of parameters $\ep^I$ $(I=1,\ldots,4)$ to depend on the spacetime coordinates $x^\mu$, since the action is invariant under the supersymmetry transformations (\ref{2SUSY4}) for constants $\ep^I$. If the equations of motion are obeyed, the right hand side of (\ref{EdS2}) must vanish; Integrating by parts, we obtain
\e
\partial^\mu j^I_\mu=0.
\ee

To derive $j^I_\mu$, let us first calculate the super-variation of the Chern-Simons term in (\ref{2LN4}):
\e\label{dcs4}
\d_\ep\cal{L}_{{\rm CS}}&=&\frac{1}{2}\ep^{\mu\nu\lambda}k_{mn}\partial_{\nu}(A^m_\mu \d_\ep A^n_\lambda)\nonumber\\
&&+\ep_A{}^\DA\g^{\mu\nu}(j^{mA}{}_\DA-j^{\prime m}{}_\DA{}^A)k_{mn}F^n_{\mu\nu}.
\ee

The variations of the kinematic terms of the Lagrangian (\ref{2LN4}) are give by
\e\label{dkm1}
-\frac{1}{2}\d_\ep(D_\mu\bar Z^A_a D^\mu Z^a_A)&=&-\partial_\mu(iD^\mu\bar Z^A_a\ep_A{}^\DA\psi^a_\DA)\nonumber\\
&&+iD^2\bar Z^A_a\ep_A{}^\DA\psi^a_\DA\nonumber\\
&&+iD^\mu\BZ^B_a\t_m{}^a{}_bZ^b_B\ep_A{}^\DA\g_\mu(j^{mA}{}_\DA-j^{\prime m}{}_\DA{}^A),
\ee
\e\label{dkm2}
-\frac{1}{2}\d_\ep(D_\mu\bar Z^\DA_\pa D^\mu Z^\pa_\DA)&=&-\partial_\mu(iD^\mu\bar Z^\DA_\pa\ep^\dag_\DA{}^A\psi^\pa_A)\nonumber\\
&&+iD^2Z^\pa_\DA\ep_A{}^\DA\psi^A_\pa\nonumber\\
&&+iD^\mu\BZ^\DB_\pa\t_m{}^\pa{}_\pb Z^\pb_\DB\ep_A{}^\DA\g_\mu(j^{mA}{}_\DA-j^{\prime m}{}_\DA{}^A),
\ee
\e\label{dpsik1}
\frac{1}{2}\d_\ep(i\bar\psi^A_\pa\g^\mu D_\mu\psi^\pa_A)
&=&-\frac{i}{2}\partial_\mu(\bar\psi^A_\pa\g^\mu \d_\ep\psi^\pa_A)\nonumber\\
&&+i\bar\psi^A_\pa\g^\mu(\d\psi)^{I\pa}_A\partial_\mu\ep^I\nonumber\\
&&
-i\ep_A{}^\DA\psi^A_\pa D^2Z^\pa_\DA +\ep_A{}^\DA\g^{\mu\nu}j^\prime_{m\DA}{}^AF^m_{\mu\nu}\nonumber\\
&&+\frac{i}{3}(\ep_A{}^\DA\g^\mu\psi^A_\pa)\t_m{}^\pa{}_\pb D_\mu(Z^\pb_\DB\mu^{\prime m\DB}{}_\DA)\nonumber\\
&&-i(\ep_A{}^\DA\g^\mu\psi^B_\pa)\t_m{}^\pa{}_\pb D_\mu(Z^\pb_\DA\mu^{ mA}{}_B)\nonumber\\
&&-(\ep_A{}^\DA\t_m{}^\pa{}_\pb\psi^\pb_B)[\bar\psi^B_\pa(j^{mA}{}_\DA-j^{\prime m}{}_\DA{}^A)]
\ee
and
\e\label{dpsik2}
\frac{1}{2}\d_\ep(i\bar\psi^\DA_a\g^\mu D_\mu\psi^a_\DA)
&=&-\frac{i}{2}\partial_\mu(\bar\psi^\DA_a\g^\mu \d_\ep\psi^a_\DA)\nonumber\\
&&+i\bar\psi^\DA_a\g^\mu(\d\psi)^{Ia}_\DA\partial_\mu\ep^I\nonumber\\
&&
-i\ep_A{}^\DA\psi_\DA^a D^2\bar Z_a^A -\ep_A{}^\DA\g^{\mu\nu}k_{mn}j^{mA}{}_\DA F^n_{\mu\nu}\nonumber\\
&&+\frac{i}{3}(\ep_A{}^\DA\g^\mu\psi_\DA^a)k_{mn}\t^n_{ab}D_\mu(Z^b_B\mu^{ mBA})\nonumber\\
&&-i(\ep_A{}^\DA\g^\mu\psi^{\DB a})k_{mn}\t^n_{ab}D_\mu(Z^{Ab}\mu^{\prime m}_{\DA\DB})\nonumber\\
&&-(\ep_A{}^\DA\t_m{}^a{}_b\psi^b_\DB)[\bar\psi^\DB_a(j^{mA}{}_\DA-j^{\prime m}{}_\DA{}^A)]
\ee
We have used the shorthand
\e
(\d\psi)^{I\pa}_A&=&-\g^\mu D_\mu
Z^\pa_\DB\s^I_A{}^\DB-\frac{1}{3}k_{mn}\t^{m\pa}{}_{\pb}Z^\pb_\DB\mu^{\prime
n\DB}{}_\DC\s^I_A{}^\DC+k_{mn}\t^{m\pa}{}_\pb Z^\pb_\DA\mu^{nB}{}_A\s^I_B{}^\DA
\ee
in (\ref{dpsik1}), and used the shorthand
\e
(\d\psi)^{Ia}_\DA=-\g^\mu D_\mu
Z^a_B\s^{I\dag}_\DA{}^B-\frac{1}{3}k_{mn}\t^{ma}{}_{b}Z^b_B\mu^{
nB}{}_C\s^{I\dag}_\DA{}^C+k_{mn}\t^{ma}{}_b Z^b_A\mu^{\prime n\DB}{}_\DA\s^{I\dag}_\DB{}^A
\ee
in (\ref{dpsik2}).

Let us now consider the Yukawa terms. To simplify the expressions, we define
\e\label{nutau}
\nu^m_{\DA\DB}&=&\t^m_{ab}\psi^a_\DA\psi^b_\DB,\nonumber\\ \nu^{\prime m}_{AB}&=&\t^m_{\pa\pb}\psi^\pa_A\psi^\pb_B,\nonumber\\
(\t^m\t^n)_{[ab]}&=&\frac{1}{2}(\t^m_{ac}\t^n{}^c{}_b-\t^m_{bc}\t^n{}^c{}_a),\nonumber\\
(\t^m\t^n)_{[\pa\pb]}&=&\frac{1}{2}(\t^m_{\pa\pc}\t^n{}^\pc{}_\pb-\t^m_{\pb\pc}\t^n{}^\pc{}_\pa).
\ee
The Yukawa terms are given by the third and fourth lines of (\ref{2LN4}). Their super variations read
\e\label{dyukawa1}
-\frac{i}{2}\d_\ep(j^m_{A\DB}j_m^{A\DB})&=&
i\big[i(\ep_A{}^\DA\psi^a_\DA)\t^m_{ab}(\psi^{\DB b}j_m{}^A{}_\DB)\nonumber\\
&&-(\g^\mu D_\mu Z^{Ab}\ep_A{}^\DA )j_m{}^B{}_\DA\t^m_{ab}Z^a_B\nonumber\\
&&-\frac{1}{4}(\ep_A{}^\DA j^m_{B\DA})C_{mnp}\mu^{nB}{}_C\mu^{pCA}
\nonumber\\
&&-\frac{1}{2}(\ep_A{}^\DA j^{mB}{}_\DB)C_{mnp}\mu^{p}{}_B{}^A\mu^{\prime n\DB}{}_\DA
\nonumber\\
&&+\frac{1}{2}(\ep_A{}^\DA j_m{}^{A}{}_\DB)(\t^m\t^n)_{[ab]}Z^{Ba}Z^b_B\mu^{\prime}_n{}^\DB{}_\DA
\big],
\ee
\e\label{dyukawa2}
-\frac{i}{2}\d_\ep(j^{\prime m}_{\DA B}j_m^{\prime\DA B})&=&
i\big[i(\ep_A{}^\DA\psi^{A\pa})\t_{m\pa\pb}(\psi^{\pb}_Bj^{\prime m}{}_\DA{}^B)\nonumber\\
&&-(\g^\mu D_\mu Z^{\pb}_\DA\ep_A{}^\DA )j^{\prime m}{}_\DB{}^A\t_{m\pa\pb}Z^{\DB\pa}\nonumber\\
&&-\frac{1}{4}(\ep_A{}^\DA j^{\prime m}{}_\DB{}^A)C_{mnp}\mu^{\prime p\DB}{}_\DC\mu^{\prime n\DC}{}_\DA
\nonumber\\
&&+\frac{1}{2}(\ep_A{}^\DA j^{\prime m}{}_\DB{}^B)C_{mnp}\mu^{\prime p\DB}{}_\DA\mu^{\prime nA}{}_B
\nonumber\\
&&+\frac{1}{2}(\ep_A{}^\DA j^{\prime m}{}_\DA{}^B)(\t_m\t_n)_{[\pa\pb]}Z^{\DC\pa}Z^\pb_\DC\mu^{nA}{}_B
\big],
\ee
\e\label{dyukawa3}
2i\d_\ep(j^m_{A\DB}j_m^{\prime\DB A})&=&
-2i\big[i(\ep_A{}^\DA\psi^a_\DA)\t^m_{ab}(\psi^{\DB b}j'_{m\DB}{}^A)\nonumber\\
&&-(\g^\mu D_\mu Z^{Ab}\ep_A{}^\DA )j'_{m\DA}{}^B\t^m_{ab}Z^a_B\nonumber\\
&&-\frac{1}{4}(\ep_A{}^\DA j^{\prime m}_{\DA B})C_{mnp}\mu^{nB}{}_C\mu^{pCA}
\nonumber\\
&&-\frac{1}{2}(\ep_A{}^\DA j^{\prime m}{}_\DB{}^B)C_{mnp}\mu^{p}{}_B{}^A\mu^{\prime n\DB}{}_\DA
\nonumber\\
&&+\frac{1}{2}(\ep_A{}^\DA j'_{m\DB}{}^{A})(\t^m\t^n)_{[ab]}Z^{Ba}Z^b_B\mu^{\prime}_n{}^\DB{}_\DA\nonumber\\
&&+i(\ep_A{}^\DA\psi^{A\pa})\t_{m\pa\pb}(\psi^\pb_Bj^{mB}{}_\DA)\nonumber\\
&&-(\g^\mu D_\mu Z^{\pb}_\DA\ep_A{}^\DA )j^{ mA}{}_\DB\t_{m\pa\pb}Z^{\DB\pa}\nonumber\\
&&-\frac{1}{4}(\ep_A{}^\DA j^{mA}{}_\DB)C_{mnp}\mu^{\prime p\DB}{}_\DC\mu^{\prime n\DC}{}_\DA
\nonumber\\
&&+\frac{1}{2}(\ep_A{}^\DA j^{mB}{}_\DB)C_{mnp}\mu^{\prime p\DB}{}_\DA\mu^{nA}{}_B
\nonumber\\
&&+\frac{1}{2}(\ep_A{}^\DA j^{mB}{}_\DA)(\t_m\t_n)_{[\pa\pb]}Z^{\DC\pa}Z^\pb_\DC\mu^{nA}{}_B
\big],
\ee
\e\label{dyukawa4}
\frac{i}{2}\d_\ep(\mu^{AB}_m\t^m_{\pa\pb}\psi^\pa_A\psi^\pb_B)&=&
-(\ep_A{}^\DA j^m_{B\DA})\nu^{\prime AB}_m\nonumber\\
&&-i(\g^\mu D_\mu Z^{\pa}_\DA\ep_A{}^\DA )\psi^\pb_B\t^m_{\pa\pb}\mu^{AB}_m\nonumber\\
&&-\frac{i}{6}(\ep_A{}^\DA j^{\prime p}_{\DB B})C_{mnp}\mu^{\prime n\DB}{}_\DA\mu^{mAB}
\nonumber\\
&&-\frac{i}{3}(\ep_A{}^\DA \psi^\pb_B)(\t^m\t^n)_{[\pb\pa]}Z^{\pa}_\DB \mu'_n{}^\DB{}_\DA\mu^{AB}_m
\nonumber\\
&&+\frac{i}{2}(\ep_A{}^\DA j^{\prime p}_{\DA B})C_{mnp}\mu^{nA}{}_C\mu^{mBC}
\nonumber\\
&&+i(\ep_A{}^\DA \psi^\pb_B)(\t^m\t^n)_{[\pb\pa]}Z^{\pa}_\DB\mu_n{}^A{}_C\mu^{BC}_m,
\ee
and
\e\label{dyukawa5}
\frac{i}{2}\d_\ep(\mu_{\DA\DB}^{\prime m}\nu^{\DA\DB}_m)&=&
-(\ep_A{}^\DA j^{\prime m}{}_\DB{}^A)\nu_{m\DA}{}^\DB\nonumber\\
&&-i(\g^\mu D_\mu Z^{Aa}\ep_A{}^\DA )\psi^{\DB b}\t_{mab}\mu_{\DA\DB}^{\prime m}\nonumber\\
&&-\frac{i}{6}(\ep_A{}^\DA j^{p}{}_{B}{}^\DB)C_{mnp}\mu^{nBA}\mu^{\prime m}_{\DA\DB}
\nonumber\\
&&-\frac{i}{3}(\ep_A{}^\DA \psi^{\DB b})(\t_m\t_n)_{[ba]}Z^a_B \mu^{nBA}\mu_{\DA\DB}^{\prime m}
\nonumber\\
&&-\frac{i}{2}(\ep_A{}^\DA j^{pA\DB})C_{mnp}\mu^{\prime n\DC}{}_\DA\mu^{\prime m}_{\DC\DB}
\nonumber\\
&&-i(\ep_A{}^\DA \psi^{\DB b})(\t_m\t_n)_{[ba]}Z^{Aa}\mu^{\prime n\DC}{}_\DA\mu^{\prime m}_{\DC\DB}.
\ee

The potential of the theory is given by the last two line of (\ref{2LN4}). Its supersymmetry transformation reads
\e\label{dpot}
\d_\ep\CL_{{\rm pot}}&=&\frac{i}{4}C_{mn}{}^p\ep_{A}{}^\DA j^{m}_{B\DA}\mu^{n(B}{}_C\mu_p{}^{A)C}\nonumber\\
&&+\frac{i}{4}C_{mnp}\ep_{A}{}^\DA j^{\prime m\DB A}\mu^{\prime n}_{\DC(\DB}\mu^{\prime p\DC}{}_{\DA)}\nonumber\\
&&+\frac{i}{2}\ep_{A}{}^\DA\psi^b_\DA(\t_m\t_n)_{[ab]}Z^{Aa}\mu^{\prime m\DB}{}_\DC\mu^{\prime n\DC}{}_\DB\nonumber\\
&&-\frac{i}{2}\ep_{A}{}^\DA\psi^{A\pb}(\t_m\t_n)_{[\pa\pb]}Z^{\pa}_\DA\mu^{ mB}{}_C\mu^{nC}{}_B\nonumber\\
&&+i\ep_{A}{}^\DA j^{'m}{}_\DB{}^A(\t_m\t_n)_{[ab]}Z^{Ba}Z^b_B\mu^{\prime n\DB}{}_\DA\nonumber\\
&&-i\ep_{A}{}^\DA j^m_{B\DA}(\t_m\t_n)_{[\pa\pb]}Z^{\DB\pa}Z^\pb_\DB\mu^{ nBA}.
\ee
In deriving (\ref{dpot}), we have used the identity
\begin{equation}\label{newid}
C_{mnp}\t^n_{cd}\t^p_{ab}=(\t_m\t_n)_{[ca]}\t^n_{bd}+(\t_m\t_n)_{[da]}\t^n_{cb}+
(\t_m\t_n)_{[db]}\t^n_{ac}+(\t_m\t_n)_{[cb]}\t^n_{ad}.
\end{equation}
Eq. (\ref{newid}) was derived \cite{Chen8} by using the identity $k_{mn}\t^m_{(ab}\t^n_{cd)}=0$.
There is a similar identity for the primed representation matrices $\t^m_{\pa\pb}$:
\begin{equation}\label{newid2}
C_{mnp}\t^n_{\pc\pd}\t^p_{\pa\pb}=(\t_m\t_n)_{[\pc\pa]}\t^n_{\pb\pd}+(\t_m\t_n)_{[\pd\pa]}\t^n_{\pc\pb}+
(\t_m\t_n)_{[\pd\pb]}\t^n_{\pa\pc}+(\t_m\t_n)_{[\pc\pb]}\t^n_{\pa\pd}.
\end{equation}

Combining every thing, we obtain the variation of the action
\e\label{EdS3}
\d_\ep S&=&\int d^3x\big[i\bar\psi^A_\pa\g^\mu(\d\psi)^{I\pa}_A\partial_\mu\ep^I
+i\bar\psi^\DA_a\g^\mu(\d\psi)^{Ia}_\DA\partial_\mu\ep^I\nonumber\\&&\quad\quad\quad+\mathcal{O}(D\mu j)+\mathcal{O}(D\mu^\prime j)+\mathcal{O}(D\mu j')+\mathcal{O}(D\mu^\prime j')\\&&\quad\quad\quad+\mathcal{O}(\nu j)+\mathcal{O}(\nu^\prime j)+\mathcal{O}(\nu j^\prime)+\mathcal{O}(\nu^\prime j^\prime)\nonumber\\&&\quad\quad\quad+\mathcal{O}(\mu\mu j)+\mathcal{O}(\mu^\prime\mu j)+\mathcal{O}(\mu\mu j^\prime)+\mathcal{O}(\mu\mu^\prime j^\prime)+\mathcal{O}(\mu^\prime\mu^\prime j)+\mathcal{O}(\mu^\prime\mu^\prime j^\prime)],\nonumber
\ee
where we have dropped the total derivative terms. We have denoted the terms containing
$D\mu j$ as $\mathcal{O}(D\mu j)$, where $D$, $\mu$, and $j$ stand for the covariant derivative,
``momentum map" and ``current" operators (see (\ref{maps})), respectively. The other terms have
 similar meanings. For instance, $\mathcal{O}(\nu j)$ stands for the terms containing the ``current"
  operator $j$ and the
  quantity $\nu$ defined by the first equation of (\ref{nutau}). 


The terms containing $D\mu j$  are given by the first term of the third line of (\ref{dkm1}), the fourth line of (\ref{dpsik2}), and the second line of (\ref{dyukawa1}):
\e\label{dmuj}
\mathcal{O}(D\mu j)&=&i(\ep_A{}^\DA\g_\mu j^{mA}{}_\DA) D^\mu\BZ^B_a\t_m{}^a{}_bZ^b_B\nonumber\\
&&+\frac{i}{3}(\ep_A{}^\DA\g^\mu\psi_\DA^a)k_{mn}\t^n_{ab}D_\mu(Z^b_B\mu^{ mBA})\nonumber\\
&&
-i(\g^\mu D_\mu Z^{Ab}\ep_A{}^\DA )j_m{}^B{}_\DA\t^m_{ab}Z^a_B.
\ee

$\mathcal{O}(D\mu^\prime j)$ are given by the first term of the third line of (\ref{dkm2}), the fifth line of (\ref{dpsik2}), the seventh line of (\ref{dyukawa3}), and the second line of (\ref{dyukawa5}):
\e
\mathcal{O}(D\mu^\prime j)&=&+iD^\mu\BZ^\DB_\pa\t_m{}^\pa{}_\pb Z^\pb_\DB(\ep_A{}^\DA\g_\mu j^{mA}{}_\DA)\nonumber\\
&&-i(\ep_A{}^\DA\g^\mu\psi^{\DB a})k_{mn}\t^n_{ab}D_\mu(Z^{Ab}\mu^{\prime m}_{\DA\DB})\nonumber\\
&&
+2i(\g^\mu D_\mu Z^{\pb}_\DA\ep_A{}^\DA )j^{ mA}{}_\DB\t_{m\pa\pb}Z^{\DB\pa}\nonumber\\
&&-i(\g^\mu D_\mu Z^{Aa}\ep_A{}^\DA )\psi^{\DB b}\t_{mab}\mu_{\DA\DB}^{\prime m}.
\ee

$\mathcal{O}(D\mu j^\prime)$ are given by the second line of (\ref{dyukawa3}), the second line of (\ref{dyukawa4}), the last term of (\ref{dkm1}), and the fifth line of (\ref{dpsik1}):
\e
\mathcal{O}(D\mu j^\prime)&=&2i(\g^\mu D_\mu Z^{Ab}\ep_A{}^\DA )j'_{m\DA}{}^B\t^m_{ab}Z^a_B\nonumber\\
&&-i(\g^\mu D_\mu Z^{\pa}_\DA\ep_A{}^\DA )\psi^\pb_B\t^m_{\pa\pb}\mu^{AB}_m\nonumber\\
&&
-iD^\mu\BZ^B_a\t_m{}^a{}_bZ^b_B(\ep_A{}^\DA\g_\mu j^{\prime m}{}_\DA{}^A)
\nonumber\\
&&-i(\ep_A{}^\DA\g^\mu\psi^B_\pa)\t_m{}^\pa{}_\pb D_\mu(Z^\pb_\DA\mu^{ mA}{}_B).
\ee

$\mathcal{O}(D\mu^\prime j^\prime)$ are given by the last term of (\ref{dkm2}), the fourth line of (\ref{dpsik1}), and the second line of (\ref{dyukawa2}):
\e
\mathcal{O}(D\mu^\prime j^\prime)&=&-i(\ep_A{}^\DA\g_\mu j^{\prime m}{}_\DA{}^A) D^\mu\BZ^\DB_\pa\t_m{}^\pa{}_\pb Z^\pb_\DB\nonumber\\
&&+\frac{i}{3}(\ep_A{}^\DA\g^\mu\psi^A_\pa)k_{mn}\t^{n\pa}{}_{\pb}D_\mu(Z^\pb_\DB\mu^{ m\DB}{}_\DA)\nonumber\\
&&
-i(\g^\mu D_\mu Z^{\pb}_\DA\ep_A{}^\DA )j'_{m\DB}{}^A\t^m_{\pa\pb}Z^{\DB\pa}.
\ee

$\mathcal{O}(\nu j)$ are given by the first term of (\ref{dyukawa1}) and the first term of the last line of (\ref{dpsik2}):
\e
\mathcal{O}(\nu j)&=&-(\ep_A{}^\DA\psi^a_\DA)\t^m_{ab}(\psi^{\DB b}j_m{}^A{}_\DB)\nonumber\\
&&
-(\ep_A{}^\DA\t_m{}^a{}_b\psi^b_\DB)(\bar\psi^\DB_aj^{mA}{}_\DA).
\ee

$\mathcal{O}(\nu^\prime j)$ are given by the first term of the last line of (\ref{dpsik1}), the sixth line of (\ref{dyukawa3}), and the first line of (\ref{dyukawa4}):
\e
\mathcal{O}(\nu^\prime j)&=&-(\ep_A{}^\DA\t_m{}^\pa{}_\pb\psi^\pb_B)(\bar\psi^B_\pa j^{mA}{}_\DA)\nonumber\\
&&+2(\ep_A{}^\DA\psi^{A\pa})\t_{m\pa\pb}(\psi^\pb_Bj^{mB}{}_\DA)
\nonumber\\
&&-(\ep_A{}^\DA j^m_{B\DA})\nu^{\prime AB}_m.
\ee

$\mathcal{O}(\nu j^\prime)$ are given by the second term of the last line of (\ref{dpsik2}), the first line of (\ref{dyukawa3}), and the first line of (\ref{dyukawa5}):
\e
\mathcal{O}(\nu j^\prime)&=&(\ep_A{}^\DA\t_m{}^a{}_b\psi^b_\DB)(\psi^\DB_aj^{\prime m}{}_\DA{}^A)\nonumber\\
&&+2(\ep_A{}^\DA\psi^a_\DA)\t^m_{ab}(\psi^{\DB b}j'_{m\DB}{}^A)
\nonumber\\
&&-(\ep_A{}^\DA j^{\prime m}{}_\DB{}^A)\nu_{m\DA}{}^\DB.
\ee

$\mathcal{O}(\nu^\prime j^\prime)$ are given by the second term of the last line of (\ref{dpsik1}) and the first line of (\ref{dyukawa2}):
\e
\mathcal{O}(\nu^\prime j^\prime)&=&(\ep_A{}^\DA\t_m{}^\pa{}_\pb\psi^\pb_B)(\bar\psi^B_\pa j^{\prime m}{}_\DA{}^A)\nonumber\\
&&-(\ep_A{}^\DA\psi^{A\pa})\t_{m\pa\pb}(\psi^{\pb}_Bj^{\prime m}{}_\DA{}^B).
\ee

$\mathcal{O}(\mu\mu j)$ are given by the first line of (\ref{dpot}), and the third line of (\ref{dyukawa1}):
\e
\mathcal{O}(\mu\mu j)&=&\frac{i}{4}C_{mn}{}^p\ep_{A}{}^\DA j^{m}_{B\DA}\mu^{n(B}{}_C\mu_p{}^{A)C}\nonumber\\
&&-\frac{i}{4}(\ep_A{}^\DA j^m_{B\DA})C_{mnp}\mu^{nB}{}_C\mu^{pCA}\nonumber\\
&=&0.
\ee

$\mathcal{O}(\mu^\prime\mu j)$ are given by the last line of (\ref{dpot}), the last two lines of (\ref{dyukawa1}), the last two lines of (\ref{dyukawa3}), and the third and fourth lines of (\ref{dyukawa5}):
\e
\mathcal{O}(\mu^\prime\mu j)&=&-i\ep_{A}{}^\DA j^m_{B\DA}(\t_m\t_n)_{[\pa\pb]}Z^{\DB\pa}Z^\pb_\DB\mu^{ nBA}\nonumber\\
&&-\frac{1}{2}(\ep_A{}^\DA j^{mB}{}_\DB)C_{mnp}\mu^{p}{}_B{}^A\mu^{\prime n\DB}{}_\DA
\nonumber\\
&&+\frac{1}{2}(\ep_A{}^\DA j_m{}^{A}{}_\DB)(\t^m\t^n)_{[ab]}Z^{Ba}Z^b_B\mu^{\prime}_n{}^\DB{}_\DA\nonumber\\
&&
-i(\ep_A{}^\DA j^{mB}{}_\DB)C_{mnp}\mu^{\prime p\DB}{}_\DA\mu^{nA}{}_B
\nonumber\\
&&-i(\ep_A{}^\DA j^{mB}{}_\DA)(\t_m\t_n)_{[\pa\pb]}Z^{\DC\pa}Z^\pb_\DC\mu^{nA}{}_B
\nonumber\\
&&-\frac{i}{6}(\ep_A{}^\DA j^{p}{}_{B}{}^\DB)C_{mnp}\mu^{nBA}\mu^{\prime m}_{\DA\DB}
\nonumber\\
&&-\frac{i}{3}(\ep_A{}^\DA \psi^{\DB b})(\t_m\t_n)_{[ba]}Z^a_B \mu^{nBA}\mu_{\DA\DB}^{\prime m}.
\ee

$\mathcal{O}(\mu\mu j^\prime)$ are given by the fourth line of (\ref{dpot}), and third line of (\ref{dyukawa3}), the last two lines of (\ref{dyukawa4}):
\e
\mathcal{O}(\mu\mu j^\prime)&=&-\frac{i}{2}\ep_{A}{}^\DA\psi^{A\pb}(\t_m\t_n)_{[\pa\pb]}Z^{\pa}_\DA\mu^{ mB}{}_C\mu^{nC}{}_B
\nonumber\\
&&+\frac{i}{2}(\ep_A{}^\DA j^{\prime m}_{\DA B})C_{mnp}\mu^{nB}{}_C\mu^{pCA}
\nonumber\\
&&+\frac{i}{2}(\ep_A{}^\DA j^{\prime p}_{\DA B})C_{mnp}\mu^{nA}{}_C\mu^{mBC}
\nonumber\\
&&+i(\ep_A{}^\DA \psi^\pb_B)(\t^m\t^n)_{[\pb\pa]}Z^{\pa}_\DB\mu_n{}^A{}_C\mu^{BC}_m.
\ee

$\mathcal{O}(\mu\mu^\prime j^\prime)$ are given by fifth line of (\ref{dpot}), the last two lines of (\ref{dyukawa2}), the forth and fifth lines of (\ref{dyukawa3}), and the third and fourth lines of (\ref{dyukawa4}):
\e
\mathcal{O}(\mu\mu^\prime j^\prime)&=&i\ep_{A}{}^\DA j^{'m}{}_\DB{}^A(\t_m\t_n)_{[ab]}Z^{Ba}Z^b_B\mu^{\prime n\DB}{}_\DA
\nonumber\\
&&+\frac{i}{2}(\ep_A{}^\DA j^{\prime m}{}_\DB{}^B)C_{mnp}\mu^{\prime p\DB}{}_\DA\mu^{\prime nA}{}_B
\nonumber\\
&&+\frac{i}{2}(\ep_A{}^\DA j^{\prime m}{}_\DA{}^B)(\t_m\t_n)_{[\pa\pb]}Z^{\DC\pa}Z^\pb_\DC\mu^{nA}{}_B
\nonumber\\&&+i(\ep_A{}^\DA j^{\prime m}{}_\DB{}^B)C_{mnp}\mu^{p}{}_B{}^A\mu^{\prime n\DB}{}_\DA
\nonumber\\
&&-i(\ep_A{}^\DA j'_{m\DB}{}^{A})(\t^m\t^n)_{[ab]}Z^{Ba}Z^b_B\mu^{\prime}_n{}^\DB{}_\DA\nonumber\\
&&-\frac{i}{6}(\ep_A{}^\DA j^{\prime p}_{\DB B})C_{mnp}\mu^{\prime n\DB}{}_\DA\mu^{mAB}
\nonumber\\
&&-\frac{i}{3}(\ep_A{}^\DA \psi^\pb_B)(\t^m\t^n)_{[\pb\pa]}Z^{\pa}_\DB \mu'_n{}^\DB{}_\DA\mu^{AB}_m.
\ee

$\mathcal{O}(\mu^\prime\mu^\prime j)$ are given by the third line of (\ref{dpot}), the eighth line of (\ref{dyukawa3}), and the last two lines of (\ref{dyukawa5}):
\e
\mathcal{O}(\mu^\prime\mu^\prime j)&=&\frac{i}{2}\ep_{A}{}^\DA\psi^b_\DA(\t_m\t_n)_{[ab]}Z^{Aa}\mu^{\prime m\DB}{}_\DC\mu^{\prime n\DC}{}_\DB
\nonumber\\
&&+\frac{i}{2}(\ep_A{}^\DA j^{mA}{}_\DB)C_{mnp}\mu^{\prime p\DB}{}_\DC\mu^{\prime n\DC}{}_\DA
\nonumber\\
&&-\frac{i}{2}(\ep_A{}^\DA j^{pA\DB})C_{mnp}\mu^{\prime n\DC}{}_\DA\mu^{\prime m}_{\DC\DB}
\nonumber\\
&&-i(\ep_A{}^\DA \psi^{\DB b})(\t_m\t_n)_{[ba]}Z^{Aa}\mu^{\prime n\DC}{}_\DA\mu^{\prime m}_{\DC\DB}.
\ee

$\mathcal{O}(\mu^\prime\mu^\prime j^\prime)$ are given by the second line of (\ref{dpot}) and the third line of (\ref{dyukawa2}):
\e\label{mu2pmupj}
\mathcal{O}(\mu^\prime\mu^\prime j^\prime)&=&\frac{i}{4}C_{mnp}\ep_{A}{}^\DA j^{\prime m\DB A}\mu^{\prime n}_{\DC(\DB}\mu^{\prime p\DC}{}_{\DA)}\nonumber\\
&&-\frac{i}{4}(\ep_A{}^\DA j^{\prime m}{}_\DB{}^A)C_{mnp}\mu^{\prime p\DB}{}_\DC\mu^{\prime n\DC}{}_\DA\nonumber
\\&=&0.
\ee

Using the identities (\ref{FI}), (\ref{newid}), (\ref{newid2}), and the identities in Appendix \ref{Identities}, it is not difficult to prove that every quantity of the last three lines of (\ref{EdS3}) vanishes, i.e. $\mathcal{O}(D\mu j)=\cdots=\mathcal{O}(\mu^\prime\mu^\prime j^\prime)=0$, or (\ref{dmuj})$=\cdots=$(\ref{mu2pmupj})$=0$. As a result, only the first line of (\ref{EdS3}) remains:
\e\label{EdS4}
\d_\ep S&=&\int d^3x\big[i\bar\psi^A_\pa\g^\mu(\d\psi)^{I\pa}_A\partial_\mu\ep^I
+i\bar\psi^\DA_a\g^\mu(\d\psi)^{Ia}_\DA\partial_\mu\ep^I].
\ee
Comparing (\ref{EdS4}) with (\ref{EdS2}), we are led to the $\CN=4$ super Poincare currents
\e\label{superpcur}
j^{I}_\mu&=&-i\bar\psi^A_\pa\g_\mu(\d\psi)^{I\pa}_A
-i\bar\psi^\DA_a\g_\mu(\d\psi)^{Ia}_\DA.
\ee

\end{document}